\documentclass[12pt]{article}
\usepackage{epsfig,amssymb}
\usepackage{latexsym}

\newcommand{\bq}{\begin{equation}}
\newcommand{\eq}{\end{equation}}
\newcommand{\bqa}{\begin{eqnarray}}
\newcommand{\eqa}{\end{eqnarray}}
\newcommand{\ben}{\begin{enumerate}}
\newcommand{\een}{\end{enumerate}}
\newcommand{\bc}{\begin{center}}
\newcommand{\ec}{\end{center}}
\newcommand{\bqb}{\begin{eqnarray*}}
\newcommand{\eqb}{\end{eqnarray*}}

\def\gsim{\gtrsim}
\def\lsim{\lesssim}

\def\pr#1#2#3{ Phys. Rev. ${\bf{#1}}$:#2 (#3)}
\def\prl#1#2#3{ Phys. Rev. Lett. ${\bf{#1}}$:#2 (#3)}

\def\prep#1#2#3{ Phys. Rep. ${\bf{#1}}$:#2 (#3)}

\def\np#1#2#3{ Nucl. Phys. ${\bf{#1}}$:#2 (#3)}

\def\epj#1#2#3{ Eur. Phys. J. ${\bf{#1}}$:#2 (#3)}
\def\ijmp#1#2#3{ Int. J. Mod. Phys. ${\bf{#1}}$:#2 (#3)}

\def\cpc#1#2#3{Comput. Phys. Commun. ${\bf{#1}}$:#2 (#3)}

\def\polon#1#2#3{Acta Phys. Polon. ${\bf{#1}}$:#2 (#3) }

\global\nulldelimiterspace = 0pt

\def\sw{s_W}

\def\swd{s^2_W}
\def\cwd{c^2_W}

\def\mwd{m_W^2}

\def\mz{m_Z}
\def\mzd{m_Z^2}

\def\tchi{\tilde \chi}

\def\mchi{m_{\tchi}}
\def\mdt{m_{\tilde d}}

\begin{document}
\pagenumbering{arabic}
\thispagestyle{empty}
\def\thefootnote{\fnsymbol{footnote}}
\setcounter{footnote}{1}

\begin{flushright}
March   2008 (corrected version)\\
PTA/08-009 \\
arXiv:0803.0813 [hep-ph]\\

 \end{flushright}
\vspace{2cm}
\begin{center}
{\Large\bf Genuine SUSY signatures from
$ug\to \tilde d \tchi_i^+$ and $ug \to  d W^+$, at high energies.}\\
 \vspace{1.5cm}
{\large G.J. Gounaris$^a$, J. Layssac$^b$,
and F.M. Renard$^b$}\\
\vspace{0.2cm}
$^a$Department of Theoretical Physics, Aristotle
University of Thessaloniki,\\
Gr-54124, Thessaloniki, Greece.\\
\vspace{0.2cm}
$^b$Laboratoire de Physique Th\'{e}orique et Astroparticules,
UMR 5207\\
Universit\'{e} Montpellier II,
 F-34095 Montpellier Cedex 5.\\
\end{center}

\vspace*{1.cm}
\begin{center}
{\bf Abstract}
\end{center}

We analyze  the quark-gluon induced process $u g\to \tilde d\tchi_i^+$,
including the one loop electroweak (EW)
effects in the minimal supersymmetric model (MSSM). This process
is dominated by $\tilde d_L$-production and is  determined by  four
helicity amplitudes,   three of which are violating helicity conservation,
and another one which respects it  and   is
logarithmically enhanced at high energy.
Combining   this $u g\to \tilde d_L\tchi_i^+$ analysis with
a  corresponding  one for $u g \to d W^+$,   we obtain simple
approximate relations between the two processes,  testing
the  MSSM structure at the amplitude  and the cross section levels.
These relations become exact at asymptotic energies
and, provided the SUSY scale is not too heavy,
 they may be approximately correct at LHC energies also.
In addition to  these, we study the 1loop EW corrections
to the inclusive $\tilde d\tchi_i^+$
production at  LHC,  presenting  as examples, the $p_T$
and angular distributions.
Comparing these   to the   corresponding
predictions for   $W$+jet production derived earlier, provides an accurate
test of the supersymmetric assignments.

\vspace{0.5cm}

\def\thefootnote{\arabic{footnote}}
\setcounter{footnote}{0}
\clearpage

\section{Introduction}

In a previous paper  we have shown  that the 1loop virtual SUSY EW effects
in the process $ug\to dW^+$, present a number of remarkable
properties \cite{ugdW-our}. Among them, is the role of SUSY in ensuring
the validity of Helicity Conservation (HC) for any two-body process at high
energy, to all orders in perturbation theory \cite{heli}.
By this we mean the fact  that at very high energies and fixed angles, the
only surviving  two-body amplitudes are those where the sum
of the initial particle helicities equal to the sum of the
final particle helicities  \cite{heli}. According to HC, these are the only
 amplitudes that could possibly contribute
at asymptotic energies, and in fact receive   the logarithmic
enhancements extensively studied in \cite{MSSMrules} and \cite{SMrules}.
All the rest must vanish in this limit.

These results raised several questions concerning
the deeper reasons for the validity of HC,
and whether  terms involving ratios
of masses could possibly violate it\footnote{Note that the general
proof in \cite{heli} is  done in the massless limit.}. Such questions
  called for further   studies of   various explicit processes \cite{ugdW-our}.
Particular among them,
 are processes involving heavy SUSY-particles in the final state,
where  establishing of HC is expected to be delayed.

Along these lines of thought, we  present here an analysis
of the process $ug\to\tilde d_L\tchi_i^+$, which starts
from the same initial state as $ug \to dW^+$,
but   its final state involves  SUSY partners of $dW^+$.
Such a study could    provide insights into  the SUSY implications,
which of course become clearest at the highest energy. \\

Denoting the helicities and momenta of the incoming
and outgoing particles in the above process as
\bq
u(p_u,\lambda_u)+g(p_g, \lambda_g) \to \tilde d_L(p_{\tilde d}) +
\tchi_i^+(p_{\tchi},\lambda_{\tchi})~~,
\label{dtchi-process}
\eq
we write  the corresponding helicity amplitudes as
$F^{\tchi}_{\lambda_u \lambda_g \lambda_{\tchi}}$.

At Born level, these  amplitudes are determined
by the two diagrams in Fig.\ref{Diag1}a, characterized
 by a $u$-quark exchange in the s-channel,
and a $\tilde d_L$-squark exchange in the  u-channel.
Because of the negligible u and d quark masses,
the charginos couple only through their pure gaugino
components, so that the appearance of a right-handed
$\tilde d_R$-squark in the final state is very strongly
suppressed. Moreover,  the incoming
u-quark must always be left-handed, with $\lambda_u=-1/2$.
The mixing in the ($\tilde d_L, \tilde d_R$)
system is also generally  negligible, since it behaves
like $m_d(A_d-\mu\tan\beta)/M_q^2$,
with  $M_q$ being  the soft SUSY breaking  mass
term. Only for extremely  large $\mu \tan\beta$
it may acquire some relevance, which can easily be taken
into account at the end, when discussing
the numerical results\footnote{See (\ref{mixing}) in Section 4.}. Therefore,
we neglect ($\tilde d_L, \tilde d_R$)-mixing in the theoretical
part of this work.

These  properties
remain true also at the 1loop EW level, as it can be seen  from the
relevant diagrams, shown  in Figs.\ref{Diag1} and \ref{Diag2}.
So,  only four independent
helicity amplitudes remain for the process (\ref{dtchi-process}),
corresponding to $\lambda_g=\pm 1$
and  $\lambda_{\tchi}=\pm 1/2$; namely
\bq
F^{\tchi}_{---}~~,~~ F^{\tchi}_{--+}~~,~~ F^{\tchi}_{-+-}~~,~~  F^{\tchi}_{-++}~~.
\label{Fchi-list}
\eq

The HC rule  would then predict that at  fixed angles, and
energies much larger than all  SUSY masses,
the first three amplitudes $F^{\tchi}_{---}$, $F^{\tchi}_{--+}$ and
$F^{\tchi}_{-+-}$   should all vanish, most often like $M^2/s$, with $M$ being
some "average" SUSY mass \cite{heli}. Only the last amplitude
$F^{\tchi}_{-++}$, which respects HC, could possibly have a non-vanishing,
  logarithmically increasing  limit \cite{heli, MSSMrules, SMrules}.

To see this explicitly, we make a complete computation  of the one loop electroweak
contributions to  the $ug\to\tilde d_L\tchi_i^+$ helicity amplitudes.
Our results  are contained in a FORTRAN code,
available at the site \cite{code}, which calculates  all four helicity
amplitudes of  (\ref{Fchi-list}), in any MSSM
with real parameters.

Using these results, we then present  the  angular and  energy dependance
of each helicity amplitude, for   three  benchmark cases covering
light, medium and high SUSY masses. This way, we try to  illustrate
how HC establishes itself at high energies. In  particular, how
    the corrections to  the leading amplitude $F^i_{-++}$
match the  high energy leading logs approximation; and how
  the   individual, relatively large contributions
to the helicity violating amplitudes, cancel each other at high energies,
and  produce   vanishing results, in accordance to the HC rule.\\

We next compare these results to those for $ug\to dW$ obtained in \cite{ugdW-our}.
Denoting the $ug\to dW$ 1loop helicity amplitudes  as
$F^W_{\lambda_u \lambda_g \lambda_d \lambda_W}$, and comparing
the leading  helicity conserving amplitudes $F^W_{----}$
and $F^W_{-+-+}$, with  the above $F^{\tchi}_{-++}$
for $\tilde d_L \tchi^+_i$-production,  we derive   relations  which
test the supersymmetric connection between the two processes at asymptotic energies.
These are subsequently transformed
to simple relations among the
differential cross sections for $ug\to\tilde d_L\tchi_i^+$ and $ug\to dW^+$,
which should be very good  asymptotically, but may be "not bad" for
LHC energies also.

Studying   experimentally these SUSY
$\sigma$-relations,   could teach us how  the asymptotic SUSY properties
are  modified  at LHC energies by   "constant" terms; i.e.  terms
 which do not depend on energy,
but may depend on the scattering angle and the SUSY masses and couplings.\\

Independently of these,  we also study   the exact  one-loop EW corrections  to
$\tilde d_L\tchi_i^+$-production at LHC, and compare it to the
corresponding study for $W$+jet production \cite{ugdW-our}.
In particular, we study the angular and transverse momentum distributions.
Such a study provides a test of whether the identification of
two "candidate particles" possibly pair-produced  at LHC,
as a $\tilde d_L$ and a $\tchi_i^+$, is consistent.  \\

The contents of the paper are: In Sect.2, the  Born
and the 1oop EW contributions to the $ug\to \tilde d_L\tchi^+_i$
helicity  amplitudes are presented, as well as  the FORTRAN code.
In Sect.3, the high energy properties of the $ug\to \tilde d_L\tchi^+_i$ and
$ug\to dW^+$ amplitudes are given, together with their
SUSY relations. The corresponding
numerical results appear in Sect.4, while in Sect.5 we give
the EW contribution to the LHC   $\tilde d_L\tchi^+_i$-production.
  Sect.6 contains  our conclusions and outlook.\\

\section{The one loop electroweak amplitudes for $u g\to\tilde d_L \tchi^+_i $}

Defining the momenta and helicities of the incoming and outgoing particles
 as  indicated in   (\ref{dtchi-process}), and using also
\bq
s=(p_u+p_g)^2 ~~,~~u=(p_{\tilde d}-p_g)^2=(p_u-p_{\tchi})^2
~~,~~ t=(p_g-p_{\tchi})^2=(p_u-p_{\tilde d})^2~~, \label{kin-stu}
\eq
we express the initial and final energies and momenta as
\bqa
 && E_u=E_g=p_u=p_g=p={\sqrt{s}\over2} ~~,~~ \nonumber \\
&& E_{\tchi}={s+m^2_{\tchi}-m^2_{\tilde d}\over2\sqrt{s}}
~~,~~E_{\tilde{d}}={s-m^2_{\tchi}+m^2_{\tilde{d}}
\over2\sqrt{s}} ~, \nonumber \\
&& p'=p_{\tilde d}=p_{\tchi}= \sqrt{E^2_{\tchi}-m^2_{\tchi}}=
\frac{\sqrt{[s-(\mchi +\mdt)^2][s-(\mchi -\mdt)^2]}}{2\sqrt{s}}~, \label{kin-Ep}
\eqa
where the mass of the  $u$-quark  has been ignored. For later use we also
give the kinematical variables
\bqa
&& R=\sqrt{E_u (E_{\tchi}+m_{\tchi})}
={1\over2}\sqrt{(\sqrt{s}+m_{\tchi})^2-m^2_{\tilde{d}}} ~, \nonumber \\
&& \beta'_{\tchi}={2p'\over\sqrt{s}}~~,~~~r_u={p\over E_u }=1
~~~,~~r_{\tchi}= {p'\over E_{\tchi}+m_{\tchi}} ~, \label{kin-Rr}
\eqa
while the c.m. scattering angle and transverse momentum are defined through
\bqa
&& u= \frac{1}{2}(\mchi^2+\mdt^2-s)
-\frac{1}{2}\sqrt{[s-(\mchi+\mdt)^2][s-(\mchi-\mdt)^2]}\cos\theta ~, \nonumber \\
&& t= \frac{1}{2}(\mchi^2+\mdt^2-s)
+\frac{1}{2}\sqrt{[s-(\mchi+\mdt)^2][s-(\mchi-\mdt)^2]}\cos\theta ~,
\label{kin-tu-theta} \\
&& \cos\theta=\sqrt{1-\frac{p_T^2}{p^{\prime 2}}} ~~~~,~~~~
|t-u|=s  \beta'_{\tchi}\sqrt{1-\frac{p_T^2}{p^{\prime 2}}} ~~.
\label{kin-theta-pT}
\eqa\\

\begin{figure}[p]
\vspace*{-1cm}
\[
\epsfig{file=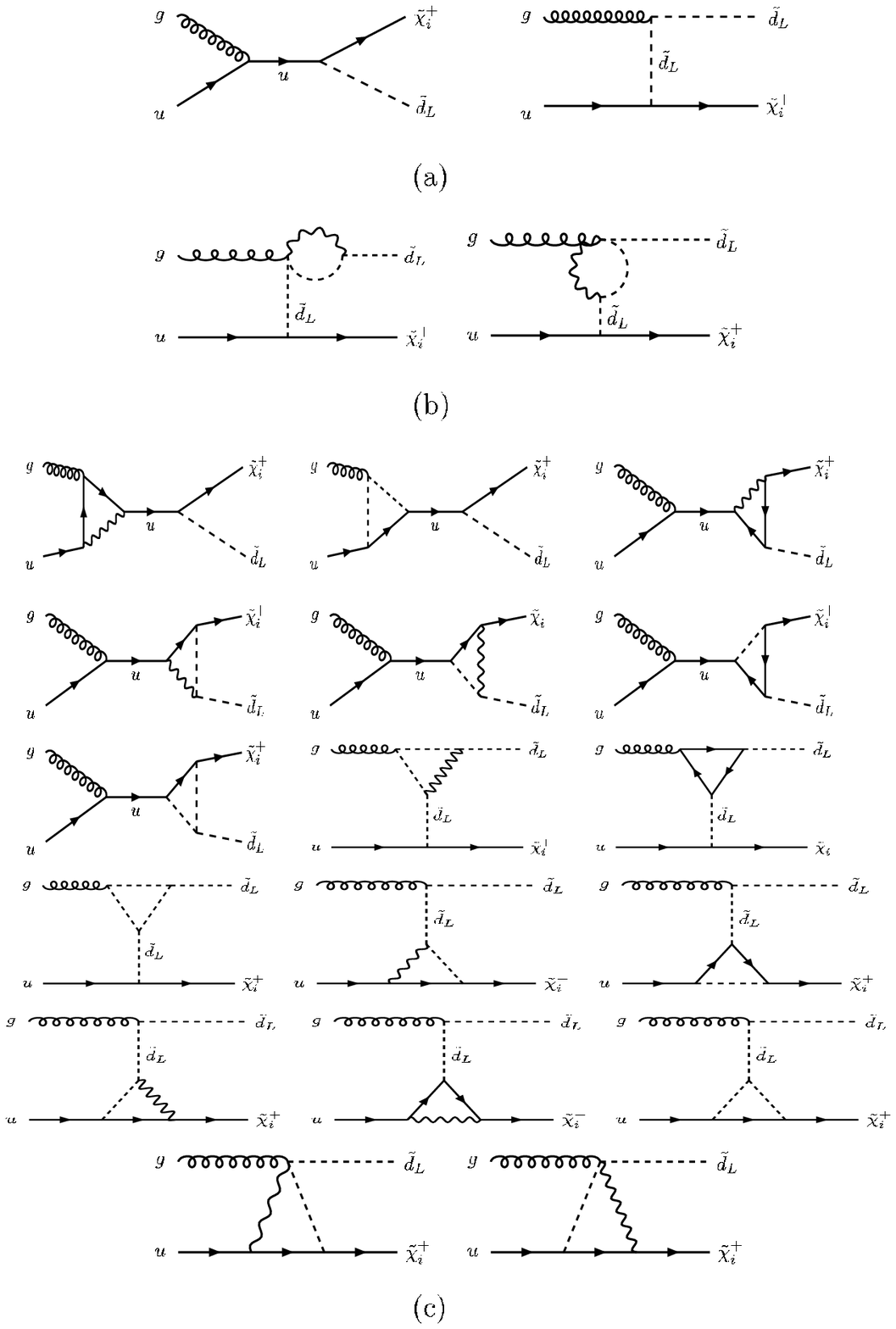,width=14cm, height=18.cm}
\]
\caption[1]{Independent Diagrams used for calculating
the $ug\to \tilde d_L \tchi_i^+$  amplitudes.
They consist of the tree diagrams (a);
the u-channel bubbles with an upper 4-leg coupling  (b);
   and the s-channel left and right triangles, together with
   the u-channel up and down triangles and the down triangles
   carrying an upper 4-leg coupling, all depicted in (c). Full, broken and
  wavy  lines describe respectively fermionic,   scalar and gauge particles. }
\label{Diag1}
\end{figure}

The Born contribution  to $u g\to\tilde d_L \tchi^+_i $,
 may  be written as
\bq
F^{\tchi, \rm ~Born}_{-\lambda_g \lambda_{\tchi}}
=F^{\tchi,~s-{\rm Born}}_{-\lambda_g \lambda_{\tchi}}
+F^{\tchi,~u-{\rm Born}}_{-\lambda_g \lambda_{\tchi}} ~~, \label{Born-Fchi}
\eq
with the two terms
\bqa
F^{\tchi,~s-{\rm Born}}_{-\lambda_g \lambda_{\tchi}}
&= & -\left ( \frac{ \lambda^a}{2}\right ) {g_sA^L_i(\tilde{d}_L)\over 2 s}
R\sqrt{2s}(1-\lambda_g)(1+2\lambda_{\tchi} r_{\tchi})
\left [\sin{\theta\over2}\delta_{\lambda_{\tchi},+}
-\cos{\theta\over2}\delta_{\lambda_{\tchi},-}\right ]~, \label{Born-Fchi-s} \\
F^{\tchi,~u-{\rm Born}}_{-\lambda_g \lambda_{\tchi}}&= &
\left ( \frac{ \lambda^a}{2}\right )
{g_sA^L_i(\tilde{d}_L)
\over u-m^2_{\tilde{d}_L}} R\sqrt{2}\lambda_g p'\sin\theta (1+2 \lambda_{\tchi} r_{\tchi})
\left [\cos{\theta\over2}\delta_{\lambda_{\tchi},+}
+\sin{\theta\over2}\delta_{\lambda_{\tchi},-}\right ] ~, \label{Born-Fchi-u}
\eqa
arising  from the two  diagrams in\footnote{We use the same conventions
as in \cite{ugdW-our}. In particular, the phase convention of the amplitude $F$ is
related to the $S$-matrix, through $S=iF$.} Fig.\ref{Diag1}a.
The   $u$-quark exchange  diagram is responsible for (\ref{Born-Fchi-s}), while
(\ref{Born-Fchi-u}) comes from $\tilde d_L$-exchange in the u-channel. The overall
factor $\lambda^a/2$ describes the color matrices acting between the initial
$u$-quark
and the final $\tilde d_L$-squark, while $g_s$ is the QCD coupling. Finally,
\bq
A^L_i(\tilde d_L)=-\frac{e}{\sw} Z^-_{1i} ~~,
\eq
expresses the $u\tilde d_L$-coupling of the  produced chargino, in terms
of its  gaugino-higgsino mixing matrix $Z^-$,
 in the notation of \cite{Rosiek}.\\

\begin{figure}[p]
\vspace*{-1cm}
\[
\epsfig{file=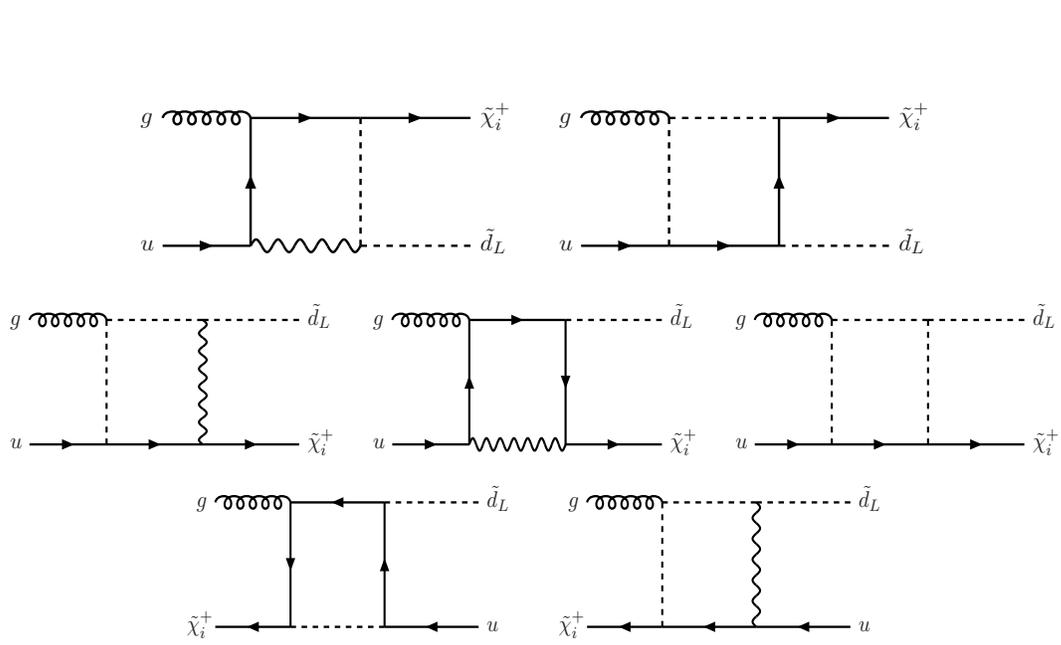,width=14cm}
\]
\caption[1]{Independent box diagrams used for $ug\to \tilde d_L \tchi_i^+$.
 Full, broken and   wavy  lines describe respectively fermionic,
 scalar and gauge particles. }
\label{Diag2}
\end{figure}

The 1loop EW corrections arise from the diagrams
in Figs.\ref{Diag1}b,c and \ref{Diag2},
to which we should also add the counter terms (c.t.) induced
by the $A^L_i(\tilde d_L)$-renormalization,   and
 the self energy (s.e.)  corrections to  the external
and internal lines of the tree diagrams in Fig.\ref{Diag1}a.\\

We first discuss  these c.t. and s.e. energy corrections, which are  simply
expressed by  modifying the Born contribution
(\ref{Born-Fchi}, \ref{Born-Fchi-s}, \ref{Born-Fchi-u}) as
\bqa
 F^{\tchi,~s-{\rm Born }}_{-\lambda_g \lambda_{\tchi}}& \to &
F^{\tchi,~s-{\rm Born}}_{-\lambda_g \lambda_{\tchi}}
\Bigg \{1+\delta Z^u_L+{1\over2} \Big [\delta Z^u_L+ \delta Z_{\tilde{d}_L}
\nonumber\\
&&+{1\over A^L_i(\tilde{d}_L)}
\sum_{j} A^L_j(\tilde{d}_L)\delta \chi^{R*}_{ji} \Big ]
+{\delta A^L_i(\tilde{d}_L)\over A^L_i(\tilde{d}_L)}
- \Big [\Sigma^u_L(s)+\delta Z^u_L \Big ] \Bigg \} ~, \label{Fchi-ct-s} \\
F^{\tchi,~u-{\rm Born }}_{-\lambda_g \lambda_{\tchi}} & \to  &
F^{\tchi,~u-{\rm Born}}_{-\lambda_g \lambda_{\tchi}}
\Bigg \{1+\delta Z_{\tilde{d}_L}+
{1\over2}\Big [\delta Z^u_L+ \delta Z_{\tilde{d}_L}
+{1\over A^L_i(\tilde{d}_L)}\sum_{j} A^L_j(\tilde{d}_L)\delta \chi^{R*}_{ji} \Big ]
\nonumber \\
&& +{\delta A^L_i(\tilde{d}_L)\over A^L_i(\tilde{d}_L)}
- \frac{\Big [\Sigma_{\tilde{d}_L}(u) -\Sigma_{\tilde{d}_L}(m^2_{\tilde{d}_L})
-(u-m^2_{\tilde{d}})\Sigma'_{\tilde{d}_L}(m^2_{\tilde{d}})
\Big ]}{u-m^2_{\tilde{d}}}
\Bigg \} ~~. \label{Fchi-ct-u}
\eqa
In the  calculation,
we always use the dimensional regularization scheme
for the ultraviolet divergencies, while
 the infrared divergencies are regularized by a "photon mass"
$m_\gamma$.

As input parameters in our renormalization scheme,
we use the $W$ and $Z$ masses,
through which the cosine of the Weinberg angle is also fixed; while the fine
structure constant  $\alpha$ is defined through the
Thompson limit  \cite{Hollikscheme}. For all couplings, we have checked
that we agree with the results of \cite{Rosiek}.

We next turn to  the various c.t. and s.e. corrections:

Defining the phase conventions
for the self energies of the  transverse gauge bosons, $u$-quark
and $\tilde d_L$-squark, so that the respective quantities $-i g^{\mu\nu}\Sigma_{VV}$,
$i\Sigma^u_L$ and $i \Sigma_{\tilde d_L}$,
always have the phase of the $S$-matrix, we find
\bqa
 && \delta Z^W_2 = - \Sigma '_{\gamma\gamma}(0)
-\frac{\alpha}{\pi}\,\frac{\cwd}{\swd}\left[ \Delta -\ln\frac{\mwd}{\mu^2} \right ]
+{c^2_W\over s^2_W}\left [{\delta m^2_Z\over m^2_Z} -
{\delta m^2_W\over m^2_W}\right ] ~~, \label{ZW}  \\
&& \delta m^2_W=Re\Sigma_{WW}(m^2_W)~~,~~\delta m^2_Z=Re\Sigma_{ZZ}(m^2_Z)
~~, \label{deltamV} \\
 && \delta Z^u_L=-\Sigma^u_L(0) ~~, \label{Zu} \\
&&  \delta Z_{\tilde{d}_L}=
- {d\Sigma_{\tilde{d}_L}(p^2)\over dp^2}\Big |_{p^2 =m^2_{\tilde{d}}}
\equiv  - \Sigma'_{\tilde{d}_L}(m^2_{\tilde{d}})  ~~. \label{Ztd}
\eqa
In all cases, these results are expressed in terms of
simple $B_j$ Passarino-Veltman (PV) functions \cite{Veltman}.
Particularly for the gauge boson s.e., the relevant results
may be obtained from the appendices\footnote{Since, as in \cite{ugdW-our},
we always regularize the infrared divergencies by a "photon mass"
$m_\gamma$, the quantity $\frac{\alpha}{2\pi} m_\gamma^2 \Delta$ must be added
to the r.h.s. of the expression (C.18) of \cite{eeV1V2-long}. }
of \cite{eeV1V2-long}.

The $A_j^L(\tilde d_L)$-dependent terms in (\ref{Fchi-ct-s}, \ref{Fchi-ct-u})
arise from the chargino  renormalization-matrices
and the $A_i^L(\tilde d_L)$-renormalization.
 Below we only present the part needed
here, following   \cite{chargino-self}.
The necessity for  $2 \otimes 2$
chargino renormalization matrices arises from the existence of two charginos,
whose mixing is affected by the 1loop  self energy bubbles. They are defined through
\bq
\tchi^+_i\to \left (\delta_{ij}
+{1\over2} [\delta\chi^L_{ij}P_L+\delta\chi^R_{ij}P_R] \right )\tchi^+_j ~.
\label{chi-renorm}
\eq
Defining then the chargino 1loop s.e. bubble contribution for the transition
 $\tchi_j^+ (p) \to \tchi_i^+(p)$, as
\bq
\Sigma_{ij}(p)=\rlap / p  P_L \Sigma^L_{ij}
+\rlap / p P_R \Sigma^R_{ij}+P_L \Sigma^S_{ij}
+P_R \Sigma^{\bar S}_{ij} ~~, \label{chargino-Sij}
\eq
with $p$ denoting the corresponding momentum,
and choosing the phase as for the other fermions\footnote{
More explicitly the phase of $ i\Sigma_{ij}$
is chosen the same as for the $S$-matrix.}, we obtain
\bq
 \delta\chi^R_{ii}=-  Re \{\Sigma^{R}_{ii}(M^2_i)
+M^2_i[\Sigma^{L\,\prime}_{ii}(M^2_i)+\Sigma^{R\prime}_{ii}(M^2_i)]
+M_i[\Sigma^{S \prime}_{ii}(M^2_i)+\Sigma^{\bar S \prime}_{ii}(M^2_i)]\}~,
\label{chiRii}
\eq
if  the initial and final charginos are  of the same kind, and
\bq
 \delta\chi^R_{ij}={2\over (M^2_i-M^2_j)}
 Re \{M^2_j\Sigma^{R}_{ij}(M^2_j)
+M_iM_j\Sigma^{L}_{ij}(M^2_j)
+M_j\Sigma^{S}_{ij}(M^2_j)+M_i\Sigma^{\bar S}_{ij}(M^2_j)\}~~, \label{chiRij}
\eq
when they are different.  Here $M_j$ denotes
the chargino masses for $j=1,2$, and we also have
$\Sigma^{\bar S}_{ij}=\Sigma^{S*}_{ji}$ \cite{chargino-self}.

The expression needed in (\ref{Fchi-ct-s}, \ref{Fchi-ct-u})
is then written as \cite{chargino-self}
\bq
\frac{\sum_{j} A^L_j(\tilde{d}_L)\delta \chi^{R*}_{ji}}
{2 A^L_i(\tilde{d}_L)}+\frac{\delta A_i^L(\tilde d_L)}{ A_i^L(\tilde d_L)}=
-\frac{\alpha}{2\pi \swd}
\left[ \Delta -\ln\frac{\mwd}{\mu^2} \right ] -{1\over2}\delta Z^W_2
+\frac{\sum_jZ^-_{1j}(\delta \chi^{R}_{ij}
+\delta \chi^{R*}_{ji})}{4 Z^-_{1i}} ~~, \label{deltaAL}
\eq
where $\Delta$ is the usual ultraviolet contribution, and $Z^-$ has been already
defined. The bubbles contributing to $\Sigma_{ij}$ consist of the exchanges
\[
(\tchi^+_k \gamma)~,~ ( \tchi^+_k Z)~,~ ( \tchi^0_kW ^+) ~,~
(\tchi^+_k H^0)~,~(\tchi^+_k h^0)~,~ (\tchi^+_k A^0)~,~ (\tchi^+_k G^0)~,~
( \tchi^0_k H^+)~,~( \tchi^0_k G^+)~,~
\]
as well as the fermion-sfermion bubbles.
They have  all been  expressed in terms of $B_j$ functions.

Using (\ref{ZW}-\ref{deltaAL}) and the substitutions
(\ref{Fchi-ct-s}, \ref{Fchi-ct-u})
in (\ref{Born-Fchi}), we obtain the full  contribution arising from
the Born terms in Fig.\ref{Diag1}a, to which  the counter terms
and self energy contributions have been inserted. All these contributions
 have the form of 1loop bubbles with two external legs.

It is worth remarking here, that inserting the s.e. and c.t. corrections
   in (\ref{Fchi-ct-s}, \ref{Fchi-ct-u}),  guarantees
that   we  never have to worry on whether our regularization
scheme preserves supersymmetry or not.
This is an important feature of our approach, which
was also used in \cite{ugdW-our}.\\

We next turn to the rest of the 1loop diagrams generically  depicted
in Figs.\ref{Diag1}b,c and \ref{Diag2}. The  full, broken
and wavy lines in these figures, describe all possible fermion,
scalar and gauge exchanges. In more detail, these exchanges are
the following:

\begin{itemize}

\item
The u-channel bubbles, with an upper 4-leg coupling
depicted in Fig.\ref{Diag1}b, involve the exchanges
\[
(\gamma \tilde d_L) ~,~ (Z \tilde d_L) ~,~ (W^- \tilde u_L)~.
\]
\item
The  first two diagrams in Fig.\ref{Diag1}c describe
s-channel left triangles involving  the exchanges
\[
(\gamma  uu) ~,~ (Z uu)~,~ (Wdd)~,~
(\tchi^0_j \tilde u_L \tilde u_L) ~, ~
(\tchi^-_j\tilde d_L \tilde d_L)~.~
\]

\item
The next 5 diagrams in Fig.\ref{Diag1}c describe
the s-channel right  triangles with  the exchanges
\bqa
&& (u \tilde d_L\gamma)~,~(u \tilde d_L Z)~,~(\gamma \tchi^+_j u)~,~
(Z \tchi^+_j u)~,~(W^+\tchi^0_j d)~,~(\tchi^+_j \gamma \tilde d_L)~,~
(\tchi^+_j Z \tilde d_L)~,~\nonumber \\
&& (\tchi^0_j W^- \tilde u_L)~,~  (\tilde u_L d \tchi^0_j)~,~
(\tchi^0_j H^-\tilde u_L)~,~(\tchi^0_j G^-\tilde u_L)~,~
(\tchi^+_j H^0\tilde d_L)~,~(\tchi^+_j h^0\tilde d_L)~. \nonumber
\eqa

\item
The next 3 diagrams in Fig.\ref{Diag1}c describe
the u-channel up triangles   with  the exchanges
\bqa
&&  (\tilde d_L \tilde d_L \gamma )~,~(\tilde d_L \tilde d_L Z)~,~
(\tilde u_L\tilde u_LW^-)~,~(dd\tchi^0_j)~,~(uu\tchi^+_j)~,~
(\tilde d_L\tilde d_L h^0)~,~(\tilde d_L \tilde d_L H^0)~,~
\nonumber \\
&& (\tilde u_L \tilde u_L H^-)~,~(\tilde u_L \tilde u_L G^-)~. \nonumber
\eqa

\item
The next 5 diagrams in Fig.\ref{Diag1}c describe
the u-channel down  triangles   with   the exchanges
\bqa
&& (\tilde d_L u \gamma)~,~(\tilde d_L u Z)~,~ (d \tilde u_L \tchi^0_j)~,~
(\gamma \tchi^+_j\tilde d_L)~,~( Z \tchi^+_j\tilde d_L)~,~
(W^+\tchi^0_j\tilde u_L)~,~  (\tchi^+_j \gamma u)~,~
\nonumber \\
&&  (\tchi^+_j Z u)~,~(\tchi^0_jW^+d)~,~
(h^0\tchi^+_j \tilde d_L)~,~(H^0\tchi^+_j \tilde d_L)~,~
(H^-\tchi^0_j \tilde u_L)~,~(G^-\tchi^0_j \tilde u_L)~. \nonumber
\eqa

\item
The last 2 diagrams in Fig.\ref{Diag1}c describe
$u$ channel triangles with an upper
4-leg coupling, with   the exchanges
\bqa
&& (\tilde d_L u \gamma )~,~ (\tilde d_L u Z)~,~
(\gamma \tchi_j\tilde d_L)~,~ (Z \tchi_j\tilde d_L)~,~
(W\tchi_j\tilde u_L)~.~
\eqa

\item
The first  2 diagrams in Fig.\ref{Diag2}, called direct boxes,
involve the exchanges
\[
(uu\tilde d_L \gamma )~,~(uu\tilde d_L Z)~,~(\tilde u_L\tilde u_L d \tchi_j^0)~.
\]

\item
The next 3 diagrams in Fig.\ref{Diag2}, called crossed  boxes,
involve the exchanges
\bqa
&& (\tilde d_L \tilde d_L \gamma \tchi_j^+)~,~(\tilde d_L \tilde d_L Z\tchi_j^+)~,~
(\tilde u_L \tilde u_L  W^- \tchi_j^0)~,~
(uu\tchi_j^+ \gamma)~,~ (uu\tchi_j^+Z)~,~ (dd\tchi_j^0 W^+)~,~ \nonumber \\
&& (\tilde d_L\tilde d_L h^0\tchi^+_j)~,~ (\tilde d_L\tilde d_L H^0\tchi^+_j)~,~
(\tilde u_L \tilde u_L H^-\tchi^0_j)~,~ (\tilde u_L \tilde u_L G^-\tchi^0_j) ~.~
\eqa

\item
And finally, the last 2 diagrams in Fig.\ref{Diag2}, called twisted   boxes,
involve the exchanges
\[
(dd\tchi_j^0\tilde u_L)~,~(\tilde d_L\tilde d_L \gamma u)~,~
(\tilde d_L\tilde d_L Z u)~.~
\]

\end{itemize}

\vspace*{0.4cm}
Using the above procedure we calculate
the four helicity amplitudes of  $ug\to \tilde d_L \tchi^+_i$
in MSSM, at the 1loop EW order. For regularizing the infrared
divergencies we choose  $m_\gamma=\mz$. The same choice was made
in \cite{MSSMrules, ugdW-our} and
has the advantage of treating $\gamma, Z$ and $W^\pm$ on the same footing
at high energies, thus preserving the $SU(2)\otimes U(1)$ symmetry  \cite{mgamma}.

Under this choice, the results for the real and imaginary parts of
the helicity  amplitudes, for any energy and scattering angle,
they   may be obtained from   a FORTRAN code available
at the site \cite{code}.
All input parameters in that code are at the electroweak scale,
and they are assumed  real. If needed, they may be calculated
from a high scale SUSY breaking model using e.g. the
 SusSpect code \cite{suspect}.

 To eliminate possible errors,  we have checked that the code   respects
the HC theorem, so that the 3 helicity violating amplitudes
$F^{\tchi}_{---}$, $F^{\tchi}_{--+}$,  $F^{\tchi}_{-+-}$
 exactly vanish asymptotically. This is   indeed a very
 efficient tool for  identifying errors.
The reason is that the helicity violating  amplitudes receive  relatively large
1loop corrections from the various  triangle and box diagrams.
Only when these are combined, they  largely cancel each other,
producing a small  result, which vanishes asymptotically.
A seemingly innocuous error can  easily destroy this cancellation.

In addition,   we have of course checked that the divergent
$\Delta$ contributions  cancel out,
both analytically and in the code.   \\

In the illustrations presented below, we select
three  constrained MSSM benchmark models covering a range for
$\tilde d_L$ and chargino masses, within the 1 TeV range.
They  are   shown  in Table 1.
\begin{table}[h]
\begin{center}
{ Table 1: Input  parameters at the grand scale,
for three  constrained MSSM benchmark models. We always have  $\mu>0$.
All dimensional parameters are in GeV. }\\
  \vspace*{0.3cm}
\begin{small}
\begin{tabular}{||c|c|c|c||}
\hline \hline
  & FLN mSP4 & $SPS1a'$ & light SUSY      \\ \hline
 $m_{1/2}$ & 137  &   250 & 50   \\
 $m_0$ & 1674  & 70 & 60   \\
 $A_0$ & 1985 & -300  & 0   \\
$\tan\beta$ & 18.6  & 10  & 10  \\
  \hline \hline
$M_{SUSY}$ & 1500  & 350  & 40  \\
  \hline \hline
\end{tabular}
 \end{small}
\end{center}
\end{table}

The first of these benchmarks
is a "heavy" scale  model we call  FLN mSP4,
which has been suggested in    \cite{Nath}
and is consistent with all present experimental
information\footnote{As is well known,
the consistency of a constrained  MSSM model often depends
on  the top mass. In the present model $m_t=170.9$ GeV has been used.
The results of the present paper though,
are not sensitive to the top mass.}.
In this model, the $\tilde d_L$ mass is   predicted  at 1.66 TeV, while
the lightest chargino lies at 98.6 GeV.
This  model has been selected to show the effects of  heavy,
but still within the LHC range, $\tilde d_L$-masses\footnote{In  our previous work
 \cite{ugdW-our}, we have used  the FP9 model in \cite{Baer},
as a "heavy scale" example. We avoid doing it here, because its  very large
 $\tilde d_L$ mass makes    its LHC production negligible.}.
 The quantity $M_{SUSY}$
 in the last line of  Table 1, gives an average of the SUSY
 masses entering the asymptotic expressions in the next section.

For the "medium" and "light" scale examples in Table 1, we use the same models
as in \cite{ugdW-our}. Thus, for  the "medium scale", we have taken the
$SPS1a'$ model of  \cite{SPA}, which is essentially consistent with
all present knowledge \cite{Nath, OBuchmueller}.
The  "light" scale model appearing
in the last column of Table 1, is
already experimentally excluded. But it is nevertheless
used here in order to indicate what would had been the picture, if the LHC energies
were much higher than all SUSY masses. \\

In Section 4, we  show   how the various
helicity amplitudes behave with energy in these examples,
and how the HC property \cite{heli},
is asymptotically established.\\

\section{Asymptotic  amplitudes and  SUSY relations. }

We next turn to the asymptotic   helicity
amplitudes, for which simple expressions
may be given.

As expected from \cite{heli}, out of the complete list of the
 $ug\to \tilde d_L \tchi_i^+ $  helicity amplitudes given in  (\ref{Fchi-list}),
only   $F^{\tchi}_{-++}$ remains
at  asymptotic energies and fixed angles;
all the rest  must vanish. Using then the asymptotic expressions for the
PV functions, taken e.g. from \cite{techpaper}, we obtain
\bqa
Re  F^{\tchi}_{-++}   &\simeq &
 -g_s \sqrt{2}   A^L_i(\tilde d_L)\sin{\theta\over2}
\Bigg \{1+{\alpha\over4\pi}{(1+26c^2_W)\over36s^2_Wc^2_W}
\Big [ 3\ln{s \over m^2_Z}-  \ln{s\over M^2_{SUSY}}
-\ln^2{s   \over m^2_Z} \Big ]
\nonumber \\
& -& {\alpha\over4\pi s^2_W} \ln^2 {s  \over m^2_W}
- {\alpha\over4\pi}\Bigg [ {(1-10c^2_W)\over36s^2_Wc^2_W}
\Big (\ln^2{-t\over m^2_Z}-\ln^2{s   \over m^2_Z} \Big ) \nonumber\\
& +& {1\over2s^2_W}\Big ( \ln^2{-u\over m^2_Z}+\ln^2{-u\over m^2_W}
-\ln^2{s   \over m^2_Z}-\ln^2{s  \over m^2_W} \Big ) \Bigg ]
\nonumber \\
&+ &\frac{\alpha}{4\pi} \left [  C^{MSSM}_{-++}(\tchi_i)+
{(1+26c^2_W)\over 72 s^2_Wc^2_W} \ln\frac{M_{SUSY}^2}{\mzd} \right ]
\Bigg \} ~, \label{ReFchimpp-LL}
\eqa
where the leading logarithmic corrections are of course
 in accordance  with the  expectations from the general analysis
of \cite{MSSMrules}. The parameter $M_{SUSY}$ in
(\ref{ReFchimpp-LL}),    appears in the last line of Table 1.

In addition to the log-corrections, we have included in (\ref{ReFchimpp-LL})
the subleading non-logarithmic correction described by the so called
"constant" contribution $C^{MSSM}_{-++}(\tchi_i)$. In principle,
$C^{MSSM}_{-++}(\tchi_i)$ can be analytically computed from the
"constant" terms in the asymptotic expansions of the PV functions
given in \cite{techpaper}. The
expressions are lengthy though, depending on all internal and external masses
and the scattering  angle. We refrain from giving them,
and only present in  Table 2 their numerical values for $\tchi_1$,
for the three models
considered, at some choices of the angles. As seen there,
the angular dependence of $C^{MSSM}_{-++}(\tchi_1)$ is mild,
for $\theta \lsim 90^o$.

\begin{table}[h]
\begin{center}
{ Table 2: Angular dependence of
$C^{MSSM}_{-++}(\tchi_1)$,\\
  for the three  MSSM  models used here.  }\\
  \vspace*{0.3cm}
\begin{small}
\begin{tabular}{||c|c|c|c||}
\hline \hline
& \multicolumn{3}{|c|}{ $C_{-++}$   } \\ \hline
  $\theta $    & FLN mSP4 & $SPS1a'$  & "light" \\ \hline
    $30^o$ & 116  & 67 & 0 \\
     $60^o$ & 123 & 73 & 6  \\
      $90^o$ &  125  & 76 & 9  \\
       $150^o$ & 147  & 98 & 31  \\
   \hline \hline
\end{tabular}
 \end{small}
\end{center}
\end{table}

As can be checked  from the code in \cite{code}, the imaginary part of the
$F^{\tchi}_{-++}$ amplitude  is also
non-vanishing asymptotically,    behaving  as
\bqa
Im  F^{\tchi}_{-++}   &\simeq &  -  \frac{\alpha g_s}{4 \swd}
\sqrt{2}   A^L_i(\tilde d_L)\sin{\theta\over2}
\left [\ln\frac{s}{\mwd}+ \ln\frac{s}{\mzd} \right ] ~~.
\eqa\\

We next turn to the corresponding asymptotic  expressions for
$ug\to dW$  studied in
 \cite{ugdW-our}. Denoting the helicity
 amplitudes for this process    as $F^W_{\lambda_u\lambda_g\lambda_d \lambda_W}$,
we find that the list of independent ones  now is  \cite{ugdW-our}
\bq
F^W_{----}~, ~  F^W_{-+-+} ~,~ F^W_{---+}~,~ F^W_{---0}~,~ F^W_{-+--}~,~
F^W_{-+-0} ~~,
\label{FW-list}
\eq
out of which, only the first two are helicity conserving
and remain non-vanishing asymptotically.
At the 1loop level of EW corrections,  they are given by
\bqa
 ReF^W_{----} &\simeq &  {eg_s\over\sqrt{2}s_W}\left ({\lambda^a\over2}\right )
{2\over\cos{\theta\over2}}
\Bigg \{1+{\alpha\over4\pi}{(1+26c^2_W)\over36s^2_Wc^2_W}
\Big [ 3\ln{s \over m^2_Z}-\eta \ln{s\over M^2_{SUSY}}
-\ln^2{s \over m^2_Z} \Big ]
\nonumber \\
& -& {\alpha\over4\pi s^2_W} \ln^2 {s \over m^2_W}
- {\alpha\over4\pi}\Bigg [ {(1-10c^2_W)\over36s^2_Wc^2_W}
\Big (\ln^2{-t\over m^2_Z}-\ln^2{s \over m^2_Z} \Big ) \nonumber\\
& +& {1\over2s^2_W}\Big ( \ln^2{-u\over m^2_Z}+\ln^2{-u\over m^2_W}
-\ln^2{s \over m^2_Z}-\ln^2{s \over m^2_W} \Big ) \Bigg ]
\nonumber \\
&+ &\frac{\alpha}{4\pi}   C^{MSSM}_{----}(W)  \Bigg \},
 \label{ReFWmmmm-LL}\\
ReF^W_{-+-+} &\simeq & {eg_s\over\sqrt{2}s_W}\left ({\lambda^a\over2}\right )
2\cos{\theta\over2} \Bigg \{1+{\alpha\over4\pi}{(1+26c^2_W)\over36s^2_Wc^2_W}
\Big [ 3\ln{s\over m^2_Z}-\eta \ln{s\over M^2_{SUSY}}
-\ln^2{s \over m^2_Z} \Big ]
\nonumber \\
&  - & {\alpha\over4\pi s^2_W}\ln^2{-s-i\epsilon \over m^2_W}
- {\alpha\over4\pi}\Bigg [{(1-10c^2_W) \over36s^2_Wc^2_W}
\Big ( \ln^2{-t\over m^2_Z}-\ln^2{s \over m^2_Z}\Big ) \nonumber\\
& +& {1\over2s^2_W}\Big ( \ln^2{-u\over m^2_Z}+\ln^2{-u\over m^2_W}
-\ln^2{s \over m^2_Z}-\ln^2{s \over m^2_W} \Big )\Bigg]
\nonumber \\
&+ &\frac{\alpha}{4\pi}   C^{MSSM}_{-+-+}(W) \Bigg \},
\label{ReFWmpmp-LL}
\eqa
in any  MSSM model, provided the  energy is
much larger than the SUSY masses \cite{ugdW-our}.

Note that
the $M_{SUSY}$ parameter in (\ref{ReFWmmmm-LL},\ref{ReFWmpmp-LL})
has been chosen the same as in  the $ug\to \tilde d_L \tchi^+_i$ case.
This is always possible, by appropriately choosing the definition of the
 subleading "constant" contributions $C^{MSSM}_{----}(W)$ and $C^{MSSM}_{-+-+}(W)$,
  in (\ref{ReFWmmmm-LL},\ref{ReFWmpmp-LL}). These
 "constants"     turn out to be rather
insensitive to the MSSM model, but  depend mildly on the
scattering angle and the helicities\footnote{In \cite{ugdW-our} we had neglected the
helicity dependence of the "constant" terms for $ug\to dW^+$.}.
They could  be analytically
calculated using \cite{techpaper}, and their numerical values
 are given    in Table 3.
\begin{table}[h]
\begin{center}
{ Table 3: Angular dependence of  $C^{MSSM}_{-\mp-\mp}(W)$\\
for the three  MSSM  models used here.  }\\
  \vspace*{0.3cm}
\begin{small}
\begin{tabular}{||c|c|c||}
\hline \hline
$\theta $    &   $C_{----}$ & $C_{-+-+}$ \\ \hline
   $30^o$  & 22 & 14  \\
    $60^o$  & 25  & 21 \\
     $90^o$  & 23 & 23  \\
      $150^o$   &29 & 45  \\
  \hline \hline
\end{tabular}
 \end{small}
\end{center}
\end{table}\\

Comparing (\ref{ReFchimpp-LL}), with (\ref{ReFWmmmm-LL}, \ref{ReFWmpmp-LL}),
we see  that  the  only differences
between  $ug \to \tilde d_L  \tchi_i^+$ and  $ug\to d W^+$
lie in
\bq
a_{\tchi W}= \frac{\alpha}{4\pi}
{(1+26c^2_W)\over 72 s^2_Wc^2_W} \ln\frac{M_{SUSY}^2}{\mzd}  ~~,
\label{a-chi}
\eq
contributing in the last line of (\ref{ReFchimpp-LL}),
and of course in the "constant" terms. Neglecting these "constant" terms, we obtain
the   {\it F-relation}:
\bq
 \cos \left ( \theta/2 \right ) F^W_{----}  \simeq
\frac{F^W_{-+-+}}{\cos \left ( \theta/2 \right )}
 \simeq   \frac{F^{\tchi}_{-++}}
 { \sin \left ( \theta/2 \right )Z^-_{1i}
(1+a_{\tchi W})}  ~~,   \label{F-relation}
\eq
which is a genuine asymptotic  SUSY prediction,
valid at the logarithmic level.
If the exact 1loop EW results are used for calculating
the amplitudes in the various parts of (\ref{F-relation}),
then violations arise which   come either from  the "constant" terms in
 (\ref{ReFchimpp-LL}, \ref{ReFWmmmm-LL}, \ref{ReFWmpmp-LL}),  or  from
 mass-suppressed   contributions to the relevant  amplitudes.
 In Section 4 we illustrate tests of the
 {\it F-relation} in the models of Table 1. \\

Remembering that
\bq
 \frac{d\hat \sigma(u g\to \tilde d_L \tchi_i^+)}
{d\cos\theta}=  \frac{\beta_{\tchi  }'}{3072 \pi s}
 \sum_{\lambda_u \lambda_g \lambda_{\tchi} }
 |F^{\tchi}_{\lambda_u \lambda_g \lambda_{\tchi}}|^2 ~~, \label{dsigma-chi-theta}
\eq
where $\beta_{\tchi }'$ is defined in (\ref{kin-Rr}), and that
\bq
 \frac{d\hat \sigma(u g\to d W^+)}{d\cos\theta}
= \frac{\beta_W'}{3072 \pi s} \sum_{\lambda_u \lambda_g \lambda_d \lambda_W }
|F^{W}_{\lambda_u \lambda_g \lambda_d \lambda_W }|^2  ~~, \label{dsigma-W-theta}
\eq
with
\bq
 \beta_W'\simeq 1-\frac{\mwd}{s}~~, \label{betapW}
\eq
we conclude that
\bq
 \frac{d\hat \sigma(u g\to d W^+)}{d\cos\theta}
 \simeq \frac{1}{R_{iW}}
 \frac{d\hat \sigma(u g\to \tilde d_L \tchi_i^+)}{d\cos\theta} ~~,
 \label{sigma-relation}
\eq
with
\bq
R_{iW} = \frac{[s-(\mchi +\mdt)^2]^{1/2}[s-(\mchi -\mdt)^2]^{1/2}}{s-\mwd}
|Z^-_{1i}|^2 \frac{(1+a_{\tchi W})^2 \sin^2 \theta }{5+2\cos\theta +\cos^2\theta}
 ~~. \label{RiW}
\eq
In deriving (\ref{sigma-relation}), we used the fact that
all helicity violating amplitudes  vanish asymptotically.
The relation (\ref{sigma-relation}) is also a genuine asymptotic
SUSY relation that we call  $\sigma $-{\it relation}.
Its  violations could come either from the "constant" terms in
 (\ref{ReFchimpp-LL}, \ref{ReFWmmmm-LL}, \ref{ReFWmpmp-LL}),
 or from  mass-suppressed contributions to  any of
 the helicity conserving or  helicity-violating amplitudes.\\

It may also be worth remarking that (\ref{F-relation}) and
(\ref{sigma-relation}, \ref{RiW}) should remain
 true even at  energies where   the 1loop approximation for
 the helicity conserving $ug\to \tilde d_L \tchi^+_i$ and $ug\to d W^+$
 could not  be adequate, provided   the SUSY scale and $a_{\tchi W}$
 remain sufficiently small; compare (\ref{a-chi}).

Considering the approximations made in deriving (\ref{F-relation}) and
(\ref{sigma-relation}),
we would naively expect that, at not-very-high energies,
 the {\it F-relation} is more accurate than the $\sigma $-{\it relation}.
We will see in the next Section, that the actual situation is opposite.
Somehow the violations induced by the
helicity-violating amplitudes in (\ref{sigma-relation}), cancel those coming
from the helicity-conserving ones, so that (\ref{sigma-relation}) becomes
quite accurate at LHC energies; at least in the  three models of Table 1. \\

At energies much larger than the $\tilde d_L$-mass and the masses of
 \underline{both} charginos, the  $u,t$
variables become independent of the final state masses, which
then simplifies (\ref{sigma-relation}) to
\bq
 {d\hat \sigma(ug\to  dW^+)\over d\cos\theta}
\simeq \left ({u^2+s^2 \over ut (1+a_{\tchi W})^2  }\right )
\frac{1}{|Z^-_{1i}|^2} {d\hat \sigma(ug\to \tilde d_L \tchi^+_i)\over d\cos\theta}
 ~~~. \label{a1-sigma-relation}
\eq
In such a case, the orthogonality of the $Z^-$ matrix may be used to  write
\bq
 {d\hat \sigma(ug\to  dW^+)\over d\cos\theta}
\simeq \left ({u^2+s^2 \over ut (1+a_{\tchi W})^2  }\right )
\sum_i {d\hat \sigma(ug\to \tilde d_L \tchi^+_i)\over d\cos\theta}
 ~~~. \label{a2-sigma-relation}
\eq \\

\section{Numerical  Expectations.}

In this Section we present the expected  behavior of
 the helicity amplitudes and SUSY relations, in the three models of
 Table 1.

Close to  threshold for the  $ug\to \tilde d_L\tchi_i^+$ process,
we would generally  expect all
 four helicity amplitudes    (\ref{Fchi-list}) to
 have comparable magnitudes; but far above threshold we should see
 the dominance of the $F^i_{-++}$ amplitude predicted by HC \cite{heli}.

The actual situation for the $SPS1a'$ model, is illustrated in
Figs.\ref{SPA-amp-fig} describing the energy and angular
dependencies of the real parts of the four amplitudes in (\ref{Fchi-list}),
for $\tchi^+_1$-production; (the imaginary parts are always smaller or  much smaller).
The results  presented in these figures are at both, the Born
and the 1loop EW level.

As seen in Fig.\ref{SPA-amp-fig}a,
where the scattering angle has been chosen at $\theta=60^o$,
the amplitudes $F^{\tchi}_{-++}$, $F^{\tchi}_{---}$, $F^{\tchi}_{-++}$
and $F^{\tchi}_{--+}$ are  comparable in magnitude,
for energies constrained by $\sqrt{s} \lsim 1.2 {\rm TeV}$;
while $F^{\tchi}_{-+-}$ is much smaller. Moreover, at such energies
the 1loop corrections are very small,
so that the Born and the 1loop results almost coincide.

The situations changes dramatically  in Fig.\ref{SPA-amp-fig}b,
in which the energy is allowed to reach the 25TeV region. There we see,
that for $\sqrt{s} \gsim 4$TeV,
the three helicity violating amplitudes $F^{\tchi}_{---}$,
$F^{\tchi}_{-++}$, $F^{\tchi}_{-+-}$
are very small and decreasing with energy,
while   the helicity conserving $F^{\tchi}_{-++}$ dominates.
Moreover, at such energies the 1loop corrections to the helicity
conserving amplitudes
 become very large, because of the large  logarithmic
 corrections in\footnote{This  means that the 1loop approximation
cannot be adequate for the actual determination of the
 helicity conserving amplitude
at very high energies. Nevertheless, the general conclusion that
this amplitude  is much larger
than all helicity violating ones, is still true \cite{heli}.}
(\ref{ReFchimpp-LL}).

In Figs.\ref{SPA-amp-fig}c and d, the angular dependence
of the helicity amplitudes are indicated at $\sqrt{s}=1$TeV and
$\sqrt{s}=4$TeV respectively. As seen there, the predominance of
$F^{\tchi}_{-++}$ against the other three amplitudes,
 is only established  at 4TeV,  provided $\theta \lsim 150^o$.
 For larger angles, an even  higher energy is needed\footnote{This is mainly
 due to the $u$-channel propagator in the right diagram in
 Fig.\ref{Diag1}a, which needs energies much larger than $\tilde d_L$, in
 order to reach the asymptotic region.}.

The same type of effects appear
also in Figs.\ref{Lig-amp-fig} based on the "light" model of Table 1;
and in Figs.\ref{Sp4-amp-fig} based on FLN mSP4 of the same Table.
The only difference is that the predominance of $F^{\tchi}_{-++}$
appears earlier for "light" and later for FLN mSP4, due to the differences
in the SUSY threshold. We note particularly that the $F^{\tchi}_{-+-}$
amplitude is always very small, at all energies.

Qualitatively similar results arise also for $\tchi^+_2$-production,
apart from the global normalization change  induced by the  replacement
$Z^-_{11} \to Z^-_{12}$ , and  the  obvious
cross section suppression induced by the   higher chargino mass.
This can be seen from Figs.\ref{SPA-amp2-fig} illustrating the
 $SPS1a'$ model case.

We also remark on the basis of the c and d parts of
Figs.\ref{SPA-amp-fig}, \ref{Lig-amp-fig} and \ref{Sp4-amp-fig},
that each helicity amplitude has its typical angular dependence.
So even in the absence of polarization measurement,
a measurement of the angular distribution could  give
 information on  the helicity structure.
Particularly at a sufficiently high energy, where
the helicity conserving amplitude dominates,
the angular distribution can be predicted.

Similar remarks  apply also for the $ug\to dW$ case,
 where there are two helicity conserving amplitudes dominating
at very high energies, with different angular dependencies;
compare Figs.4, 7 and 10 of \cite{ugdW-our}.\\

We next turn to testing  the {\it F-relation}  (\ref{F-relation}),
at the level of our 1loop EW results.
To this aim we compare  in  Fig.\ref{F-relation-fig}a,
the 4 quantities
\bq
 \cos \left ( \theta/2 \right ) F^{dW^+}_{----} ~~,~~
\frac{F^{dW^+}_{-+-+}}{\cos \left ( \theta/2 \right )}
~~,~~   \frac{F^{\tilde d \tchi^+_1}_{-++}}{ \sin \left ( \theta/2 \right )Z^-_{11}
(1+a_{\tchi})}
   ~~,~~    \frac{F^{\tilde d \tchi^+_2}_{-++}}{ \sin \left ( \theta/2 \right )
   Z^-_{12} (1+a_{\tchi})}
   ~~, ~~  \label{test-F-relation}
\eq
  as functions of the energy, using  the "light" model and
  fixing the angle at  $\theta=60^o$.
The last two terms in (\ref{test-F-relation}) come from
$\tchi_1^+$ and $\tchi_2^+$ respectively.
Similar results are expected for other angles also.
In Figs.\ref{F-relation-fig}b,c the corresponding results for the $SPS1a'$ and
FLN mSP4 models are also shown.
As seen in these  figures, the  parts of (\ref{test-F-relation})
referring to $ud\to dW$, almost coincide at all energies, for all three models.
The deviations of the right parts though,
coming from $ug\to \tilde d_L\tchi^+_1$ or $ug\to \tilde d_L\tchi^+_2$,
depend on the scale of the MSSM model; they are  negligible for the "light" model,
and increase as we move to $SPS1a'$ and then to FLN mSP4. We note that
the relative magnitudes of these deviations become constant at high energies,
since they arise from  the  "constant" terms in (\ref{ReFchimpp-LL}) and
(\ref{ReFWmmmm-LL}, \ref{ReFWmpmp-LL}). \\

Correspondingly, the testing of  the  $\sigma$-{\it relation}  (\ref{sigma-relation}),
is done in Figs.\ref{Signa-e-relation-fig} and \ref{Signa-a-relation-fig},
at the level of our 1loop EW results.
More explicitly, we compare  in Figs.\ref{Signa-e-relation-fig}a,b,c,
the three  quantities
\bq
 \frac{d\hat \sigma(u g\to d W^+)}{d\cos\theta} ~~,~~
  \frac{1}{R_{1W}}  \frac{d \hat \sigma(u g\to \tilde d_L \tchi_1^+)}{d\cos\theta}
  ~~,~~
  \frac{1}{R_{2W}}  \frac{d \hat \sigma(u g\to \tilde d_L \tchi_2^+)}{d\cos\theta}
  ~~, \label{test-sigma-relation}
\eq
as functions of the energy, for the "light", $SPS1a'$ and FLN mSP4 models
respectively, using  $\theta=60^o$.  Correspondingly,
 in Figs.\ref{Signa-a-relation-fig}a,b,c,
we compare the angular dependencies
of the same  quantities, fixing the energy at 3 TeV.
As seen there, the $\sigma$-{\it relation} is almost perfect
for the "light" model, gradually  worsening as we move towards models with higher
supersymmetric masses; i.e. to $SPS1a'$ first, and then to FLN mSP4.
In fact this worsening is very small for $\tchi_1^+$ production,
and increases for $\tchi_2^+$ production, obviously due to the higher chargino
mass.

We may also add here that in case the ($\tilde d_L,\tilde d_R$)-mixing
is not fully negligible, and some $\tilde d_1, \tilde d_2 $ are the
true sdown squarks, then this mixing can  easily be taken into account
 by replacing in (\ref{test-sigma-relation})
\bq
\sigma(\tilde d_L) \to \frac{ \sigma(\tilde d_1)}{\cos^2\tilde \theta_d }
\simeq  \frac{ \sigma(\tilde d_2)}{\sin^2\tilde \theta_d } ~~. \label{mixing}
\eq

Comparing  Figs.\ref{F-relation-fig}, with Figs.\ref{Signa-e-relation-fig}
and \ref{Signa-a-relation-fig}, we conclude (with some surprise),
that the  $\sigma$-{\it relation}  is more accurate than the {\it F-relation}.
This is most impressive  in  the  $\tchi_1^+$ case for  the medium
and heavy scale models $SPS1a'$ and FLN mSP4,  where the low
energy {\it F-relation} deviations in Fig.\ref{F-relation-fig}b,c, are
cured in  the $\sigma$-{\it relation} Figs.\ref{Signa-a-relation-fig}b,c,
 by  contribution from the helicity violating amplitudes.
Is there a deeper reason for this? Or, it is an accidental result?
Further studies with other processes may help  clarifying this.

\section{Predictions for  $\tilde d_L \tchi^+_i$ distributions at LHC.}

Contrary to the results in the previous  Sections 3 and 4,
this  Section does not involve any asymptotic energy assumption.
Instead, the  code for the   $ug\to \tilde d_L\tchi_i^+$
helicity amplitudes presented  above, is   used  to calculate
the $\tilde d_L \tchi^+_i$ production at the actual LHC energies.

We present results, both at the Born level,
as well as at the level of the 1loop EW corrections. Our aim is to
see whether $\tilde d_L \tchi^+_i$-production and its
 SUSY 1loop corrections, are   visible at LHC.

As already said, the infrared divergencies are avoided by choosing
$m_\gamma=\mz$. All other infrared sensitive contributions,
are supposed to be included
in the pure QED contribution, following the same philosophy
as in \cite{ugdW-our}.\\

Next, we first discuss  the angular distribution in the
c.m. of the $\tilde d_L\tchi_i^+$-subprocess at LHC.
In analogy to $W$+jet production in \cite{ugdW-our}, and folding in
the needed parton distributions \cite{Durham}, this is given by
\bq
\frac{d\sigma(pp\to \tilde d_L \tchi_i^+~...)}{ds d\cos\theta }=\frac{1}{S}
\int^1_{\frac{s}{S}} \frac{dx_a}{x_a}
\left [P^i_{ang}\left (x_a,\frac{s}{Sx_a},\theta \right ) +
\tilde P^i_{ang}\left (x_a,\frac{s}{Sx_a},\theta \right )\right ] ~~,
\label{chi-ang1}
\eq
where  $s=x_ax_b S$, with $\sqrt{S}=14$ TeV being the  e LHC c.m. energy,
and
\bqa
&& P^i_{ang}\left (x_a,x_b ,\theta \right )=
{d \hat \sigma(ug\to \tilde d_L \tchi^+_i)\over d\cos\theta }
f_u(x_a)f_g(x_b) ~~, \nonumber \\
&& \tilde P^i_{ang}(x_a, x_b,\theta )=P^i_{ang}(x_b, x_a,\pi-\theta) ~~.
\label{chi-ang2}
\eqa
Here,  (\ref{dsigma-chi-theta}) should be used, and we should also remark that
the CKM-matrix effects are  negligible in (\ref{chi-ang1}).

The implied angular distributions  at  the Born
and the 1loop EW approximation, are then given
in Figs.\ref{LHC-fig}a and b, corresponding  to $\sqrt{s}=3$TeV and
$\sqrt{s}=6$TeV respectively, for the three MSSM models of Table 1.
As seen there, the overall magnitude of the cross section is at the level
of $0.1 {\rm fb/TeV^2}$ for $\sqrt{s}\gsim 3$TeV,
while the 1loop EW contribution
always reduces the Born result.
For $\sqrt{s}=3$TeV and $\theta \sim 50^o$,  this reduction is at the 30\%
level for the "light" model, the 20\% level for $SPS1a'$, and the
10\% level for FLN mSp4. Such cross sections seem difficult to measure at LHC,
mainly because of the large value of $\sqrt{s}$. Only closer to
threshold, we could get measurable
cross sections\footnote{This is elucidated by the $p_T$-discussion below.}. \\

Correspondingly, the $\tilde d_L$ or $\tchi_i^+$
transverse momentum  $(p_T)$ distribution at LHC is  determined  by
first noting that
\bq
 \frac{d\hat \sigma(u g\to \tilde d_L \tchi_i^+)}{d p_T}=
 \frac{p_T }{768 \pi s|t-u|}
\left [ \sum_{\lambda_u \lambda_g \lambda_{\tchi} }
 |F^{\tchi}_{\lambda_u \lambda_g \lambda_{\tchi}}|^2 \Big |_\theta
 +\sum_{\lambda_u \lambda_g \lambda_{\tchi} }
 |F^{\tchi}_{\lambda_u \lambda_g \lambda_{\tchi}}|^2 \Big |_{\pi -\theta}
 \right ]~~, \label{dsigma-chi-pT}
\eq
where (\ref{kin-theta-pT}) is used, and then using
\bq
\frac{d\sigma(pp\to \tilde d_L \tchi_i^+~...)}{dp_T}=
\int^1_0 dx_a\int^1_0 dx_b
\theta(x_ax_b-\tau_m) [P^i_T(x_a,x_b) +\tilde P^i_T(x_a,x_b)]~~,
\label{chi-pT1}
\eq
where
\bqa
&& \tau_m={1\over S}\left (\sqrt{p^2_T+m_{\tchi }^2}
+\sqrt{p^2_T+m_{\tilde d}^2}\right )^2 ~~, \label{kin-LHC1} \\
 && P^i_T(x_a, x_b)={d \hat \sigma(ug\to \tilde d_L \tchi^+_i)\over dp_T}
f_u(x_a)f_g(x_b) ~~~, \nonumber \\
&&  \tilde P^i_T(x_a, x_b)=P^i_T(x_b, x_a) ~~. \label{chi-pT2}
\eqa
The relevant results for the three models in Table 1 are presented in
Fig.\ref{LHC-fig}c, again  for the Born predictions
and the 1loop EW corrections. As before, the 1loop contribution
 always reduces the Born prediction, by an amount which
 for $p_T\sim 0.6$TeV lies at the level of 18\% for the "light" model,
 11\% for $SPS1a'$, and 7\% for FLN mSP4.  For $W$+jet production, the
 corresponding effect was found at the 10\% level \cite{ugdW-our}.

 For an integrated LHC luminosity of 10 or 100 ${\rm fb^{-1}/TeV}$,
it   seems possible to measure this direct
$\tilde d_L \tchi^+_i$ production, assuming that the masses are not too high.
To achieve this, the experiment
needs of course to include properly all necessary infrared QED, and the
higher order QCD effects.

 The ratio of the $\tilde d_L\tchi_i^+$  LHC distributions,
given (\ref{chi-ang1}) and (\ref{chi-pT1}), with respect to the corresponding
quantities for $W$+jet production studied in \cite{ugdW-our},
 may  then provide a basic   test of
the supersymmetric nature. For sufficiently light  SUSY masses,
it   may even be possible to determine  the  1loop EW reductions of the Born
contributions. \\

\section{Summary and Conclusions}

In this paper we have calculated the four independent  helicity amplitudes for
the process $ug\to \tilde d_L \tchi_i^+$,
to 1loop EW order in MSSM.  The  results
are contained in a   code, valid for any set
of real MSSM parameters in the EW scale, and released at \cite{code}.

Combining these results,  with the previous ones in
\cite{ugdW-our}, we pursued the following  three  aims.

\begin{figure}[p]
\vspace*{-2.cm}
\[
\hspace{-0.5cm}
\epsfig{file=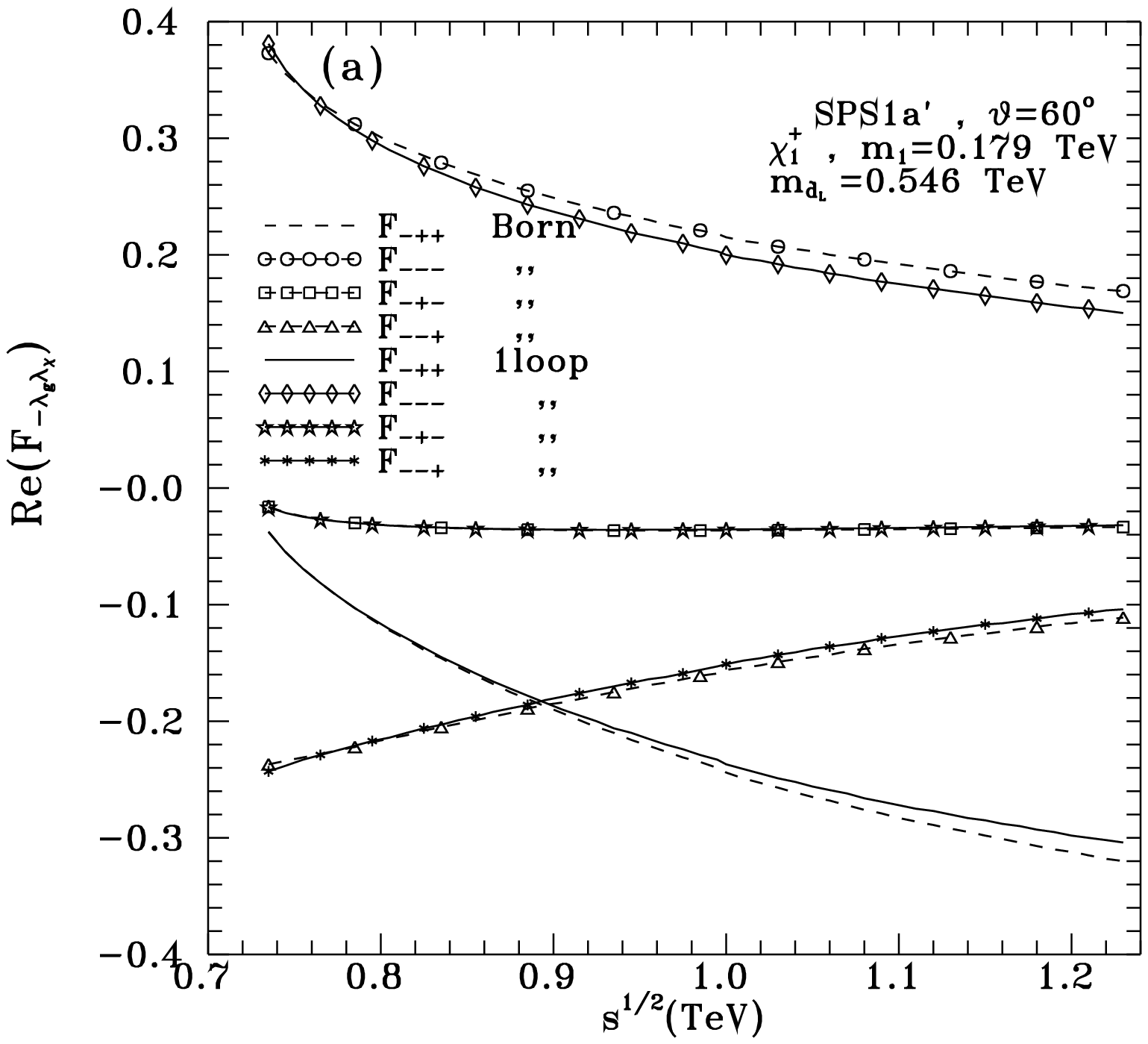,height=7.5cm}\hspace{0.5cm}
\epsfig{file=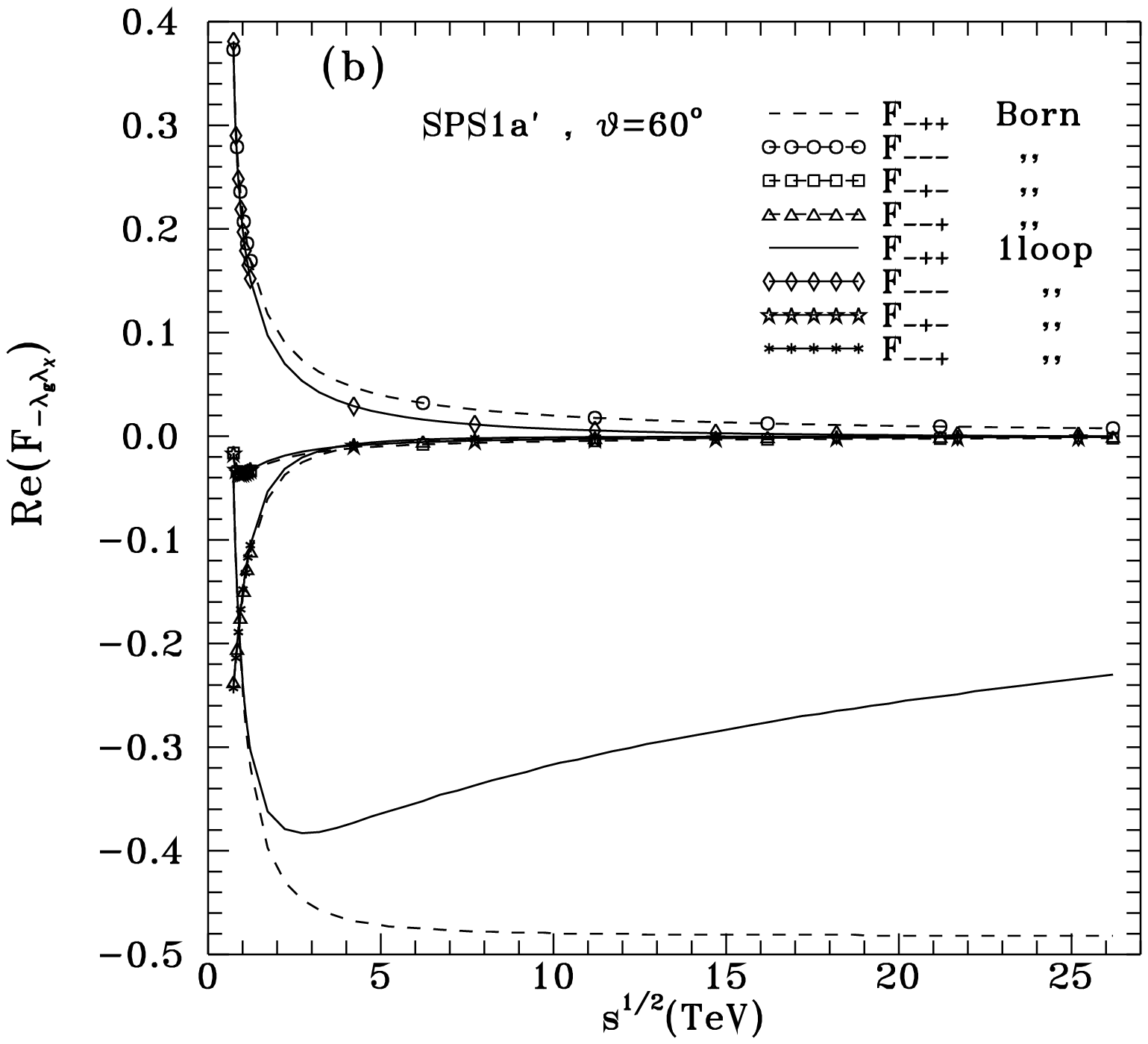,height=7.5cm}
\]
\vspace*{0.5cm}
\[
\hspace{-0.5cm}
\epsfig{file=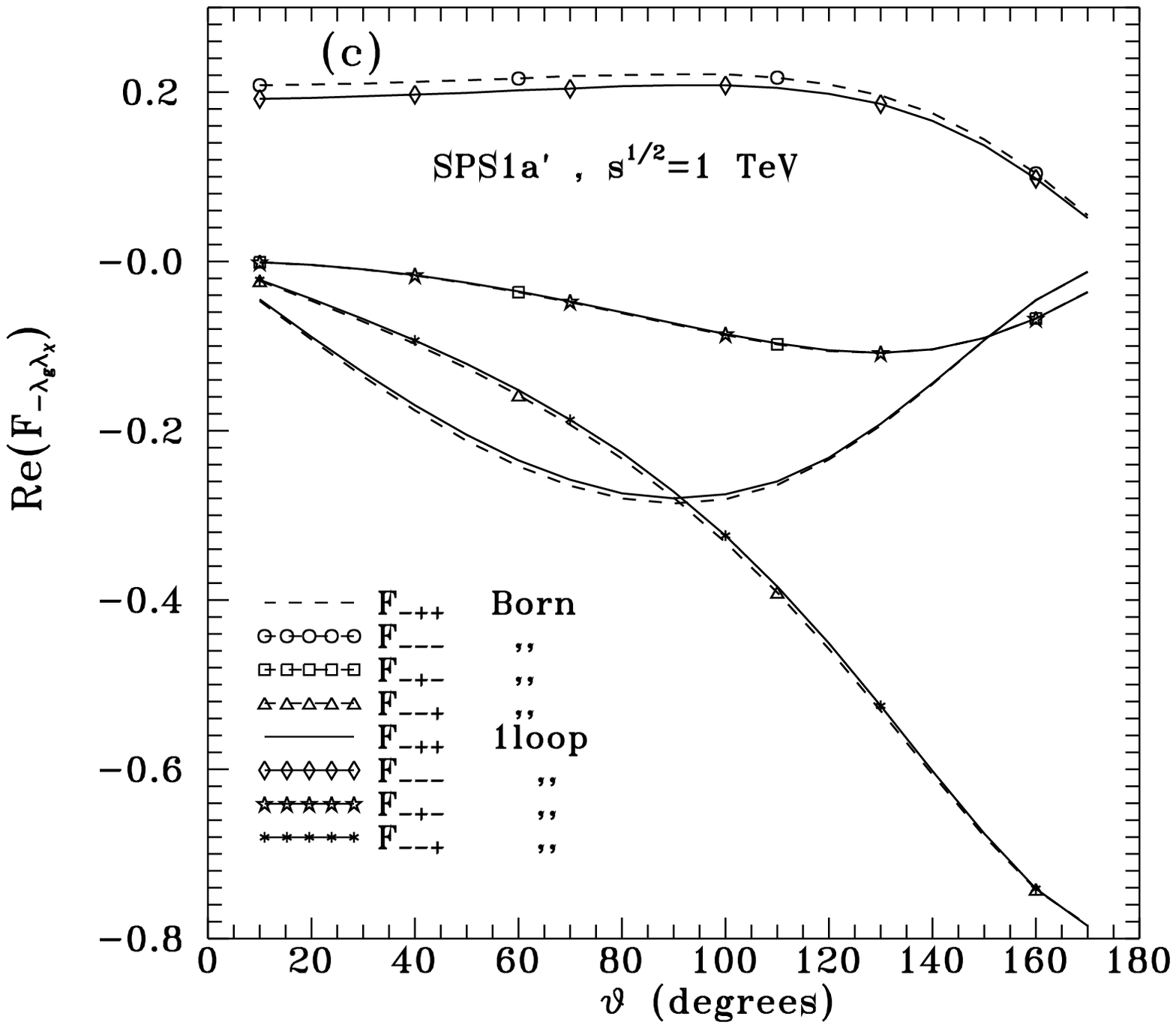,height=7.5cm}\hspace{0.5cm}
\epsfig{file=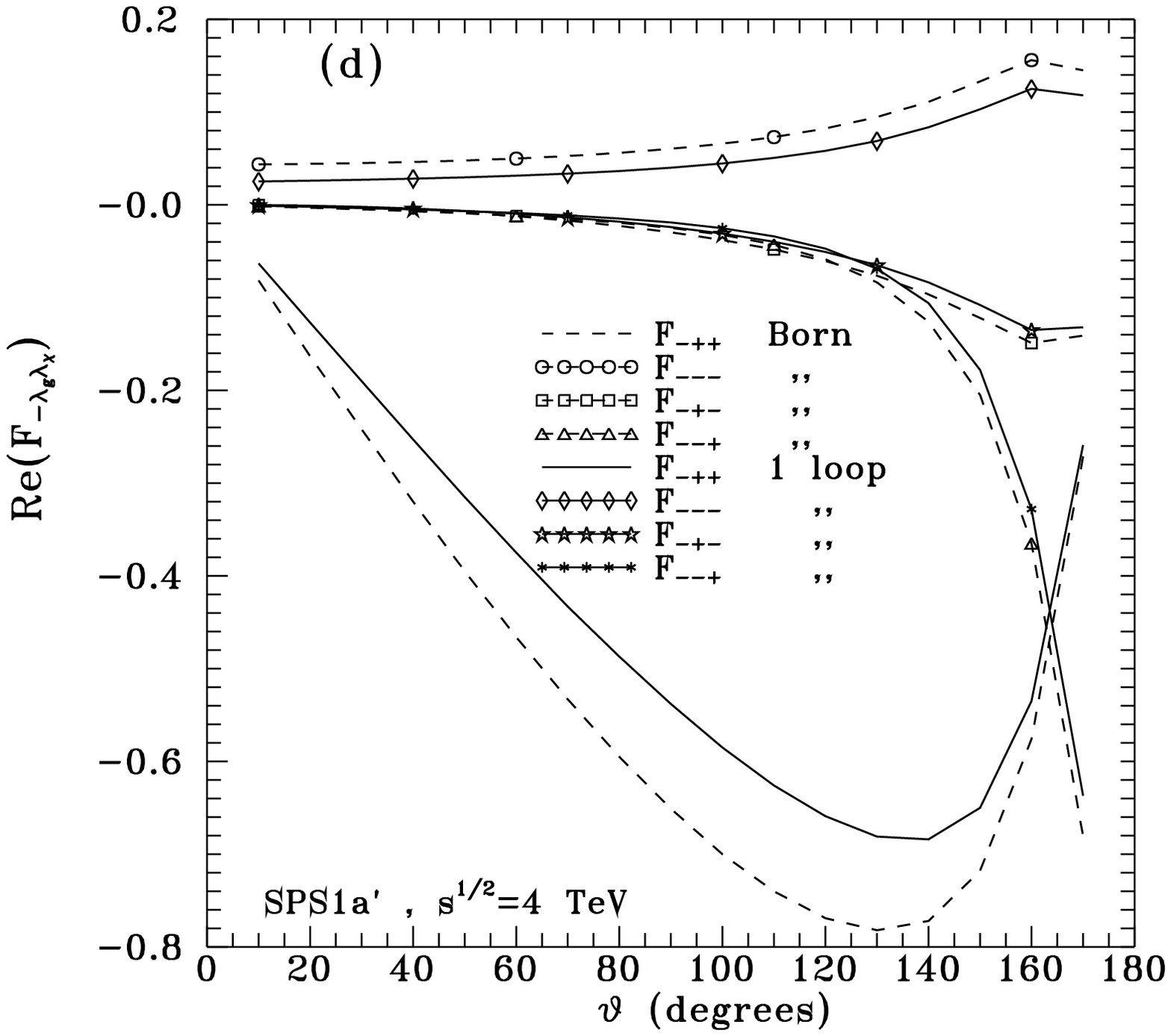,height=7.5cm}
\]
\caption[1]{The  $ug\to \tilde d_L \tchi^+_1$ helicity amplitudes
at $\theta=60^o$, for the $SPS1a'$ model. The  energy dependencies
cover an LHC-type range (a), and a higher energy region (b);
while the  angular dependencies are given
at $\sqrt{s}=1$TeV (c), and $\sqrt{s}=4$TeV (d). }
\label{SPA-amp-fig}
\end{figure}

\begin{figure}[p]
\vspace*{-2.cm}
\[
\hspace{-0.5cm}
\epsfig{file=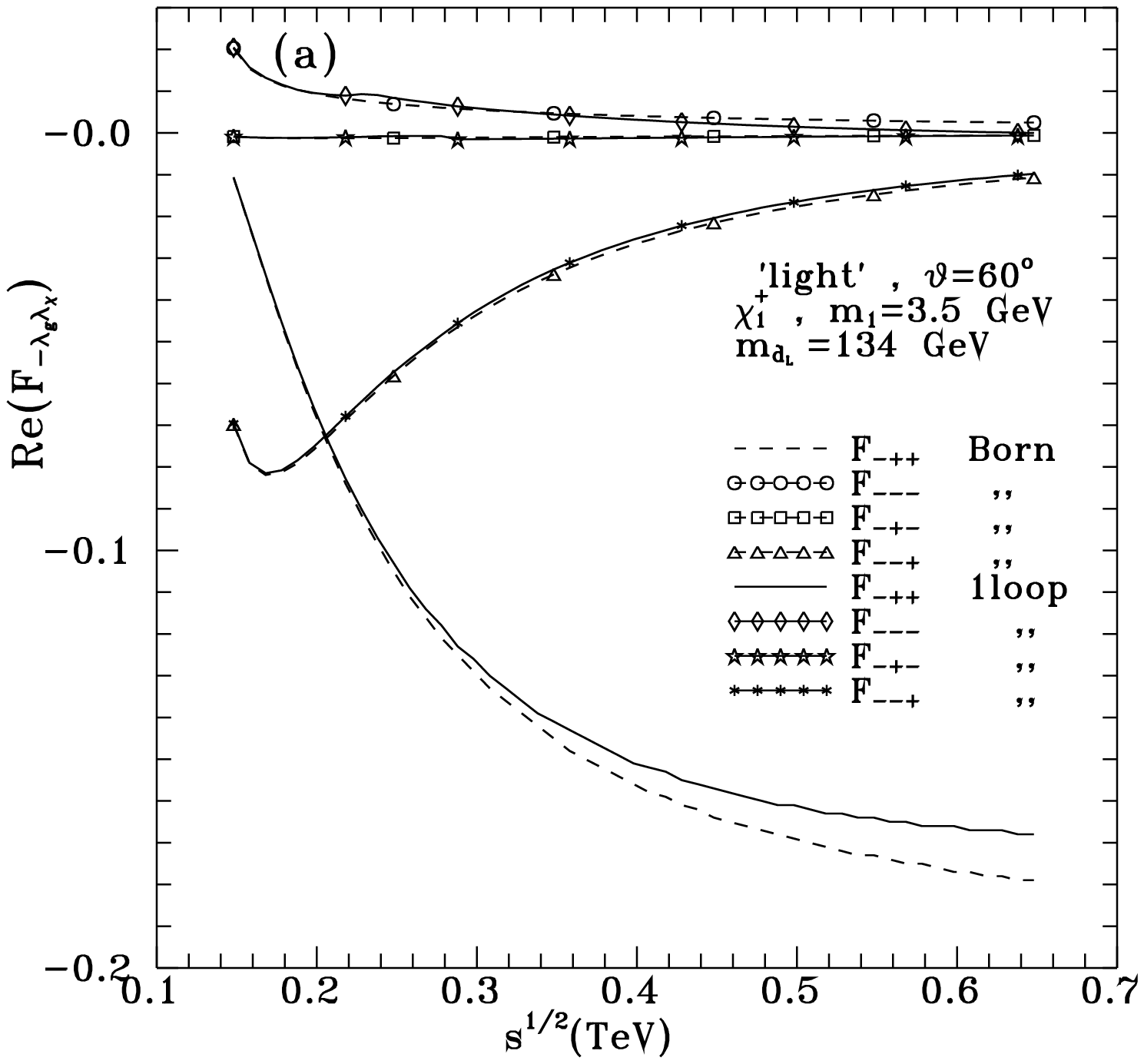,height=7.5cm}\hspace{0.7cm}
\epsfig{file=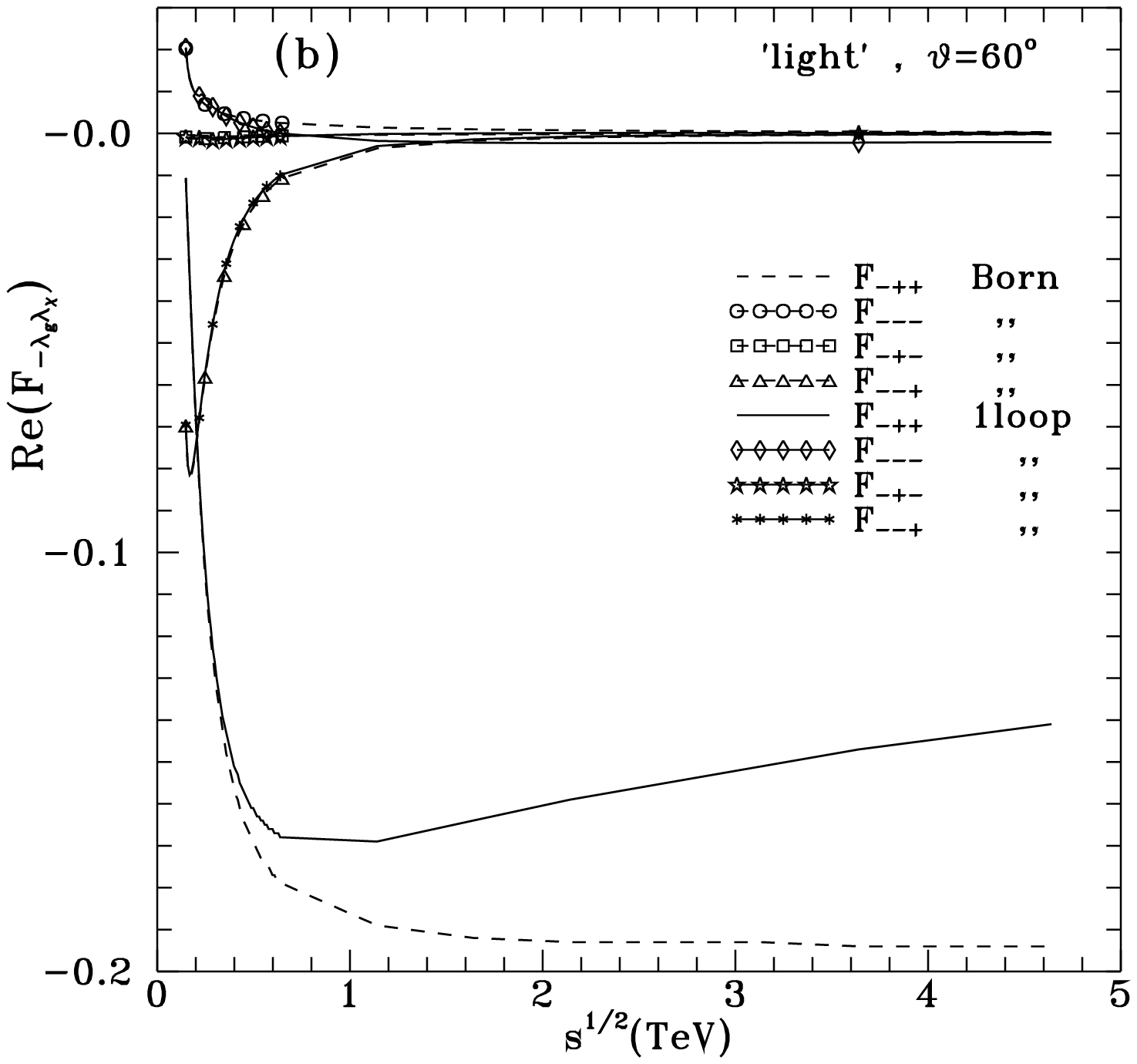,height=7.5cm}
\]
\vspace*{0.5cm}
\[
\hspace{-1.cm}
\epsfig{file=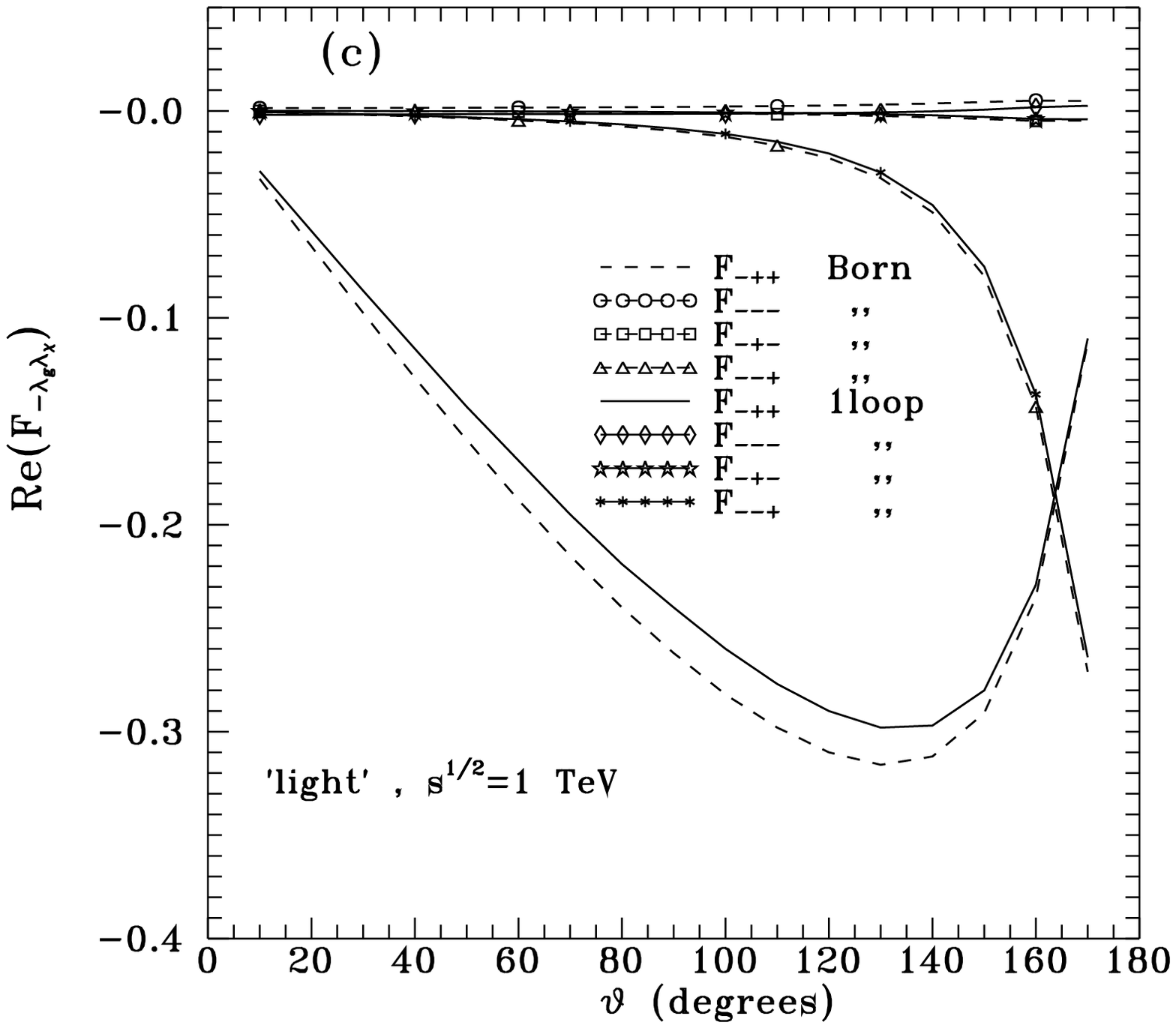,height=7.5cm}\hspace{0.5cm}
\epsfig{file=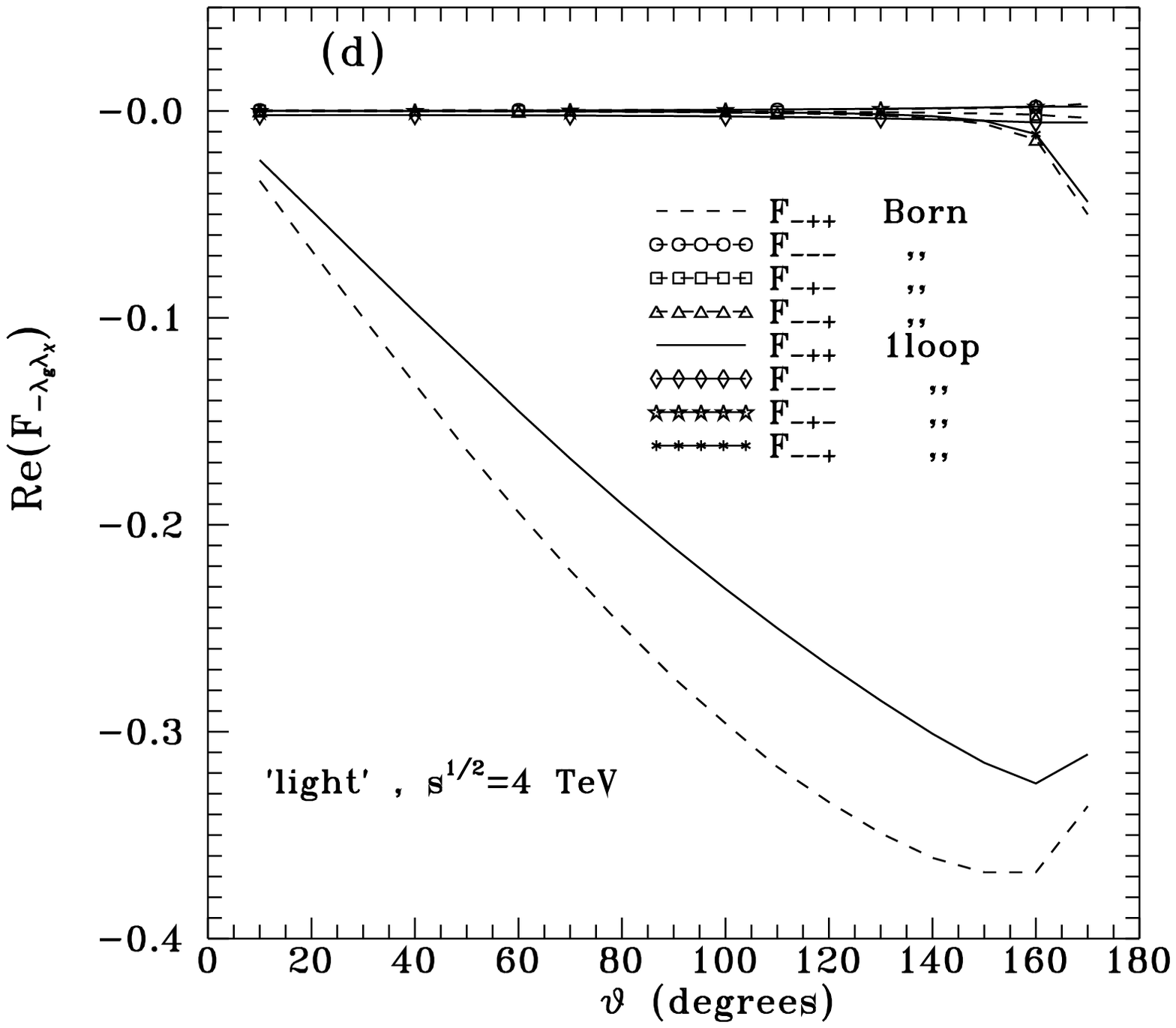,height=7.5cm}
\]
\caption[1]{The  $ug\to \tilde d_L \tchi^+_1$ helicity amplitudes
for the "light" model. The  energy dependencies
cover an LHC-type range (a), and a higher energy region (b);
while the  angular dependencies are given
at $\sqrt{s}=1$TeV (c), and $\sqrt{s}=4$TeV (d). }
\label{Lig-amp-fig}
\end{figure}

\begin{figure}[p]
\vspace*{-2.cm}
\[
\hspace{-0.5cm}
\epsfig{file=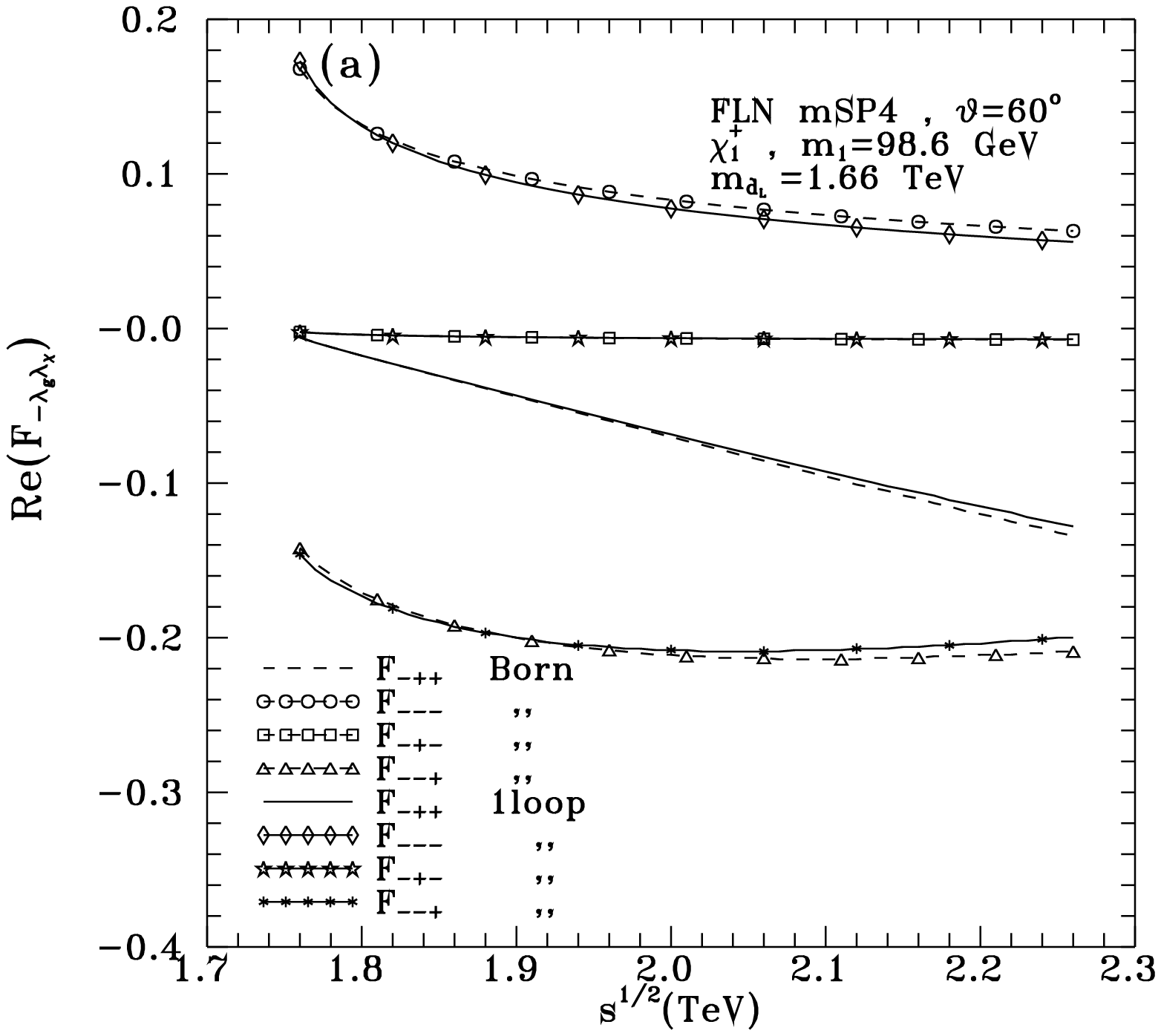,height=7.5cm}\hspace{0.5cm}
\epsfig{file=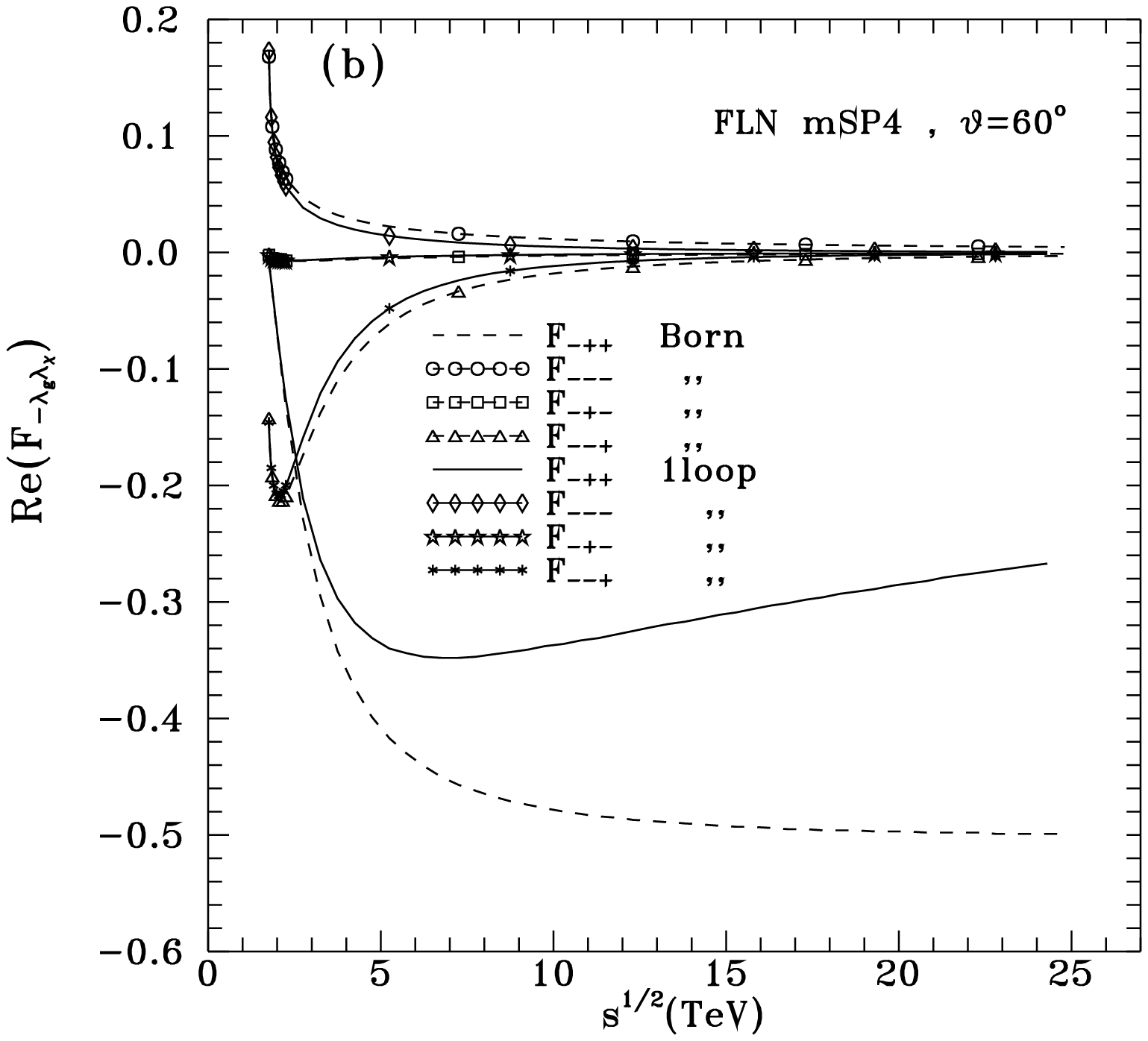,height=7.5cm}
\]
\vspace*{0.5cm}
\[
\hspace{-0.5cm}
\epsfig{file=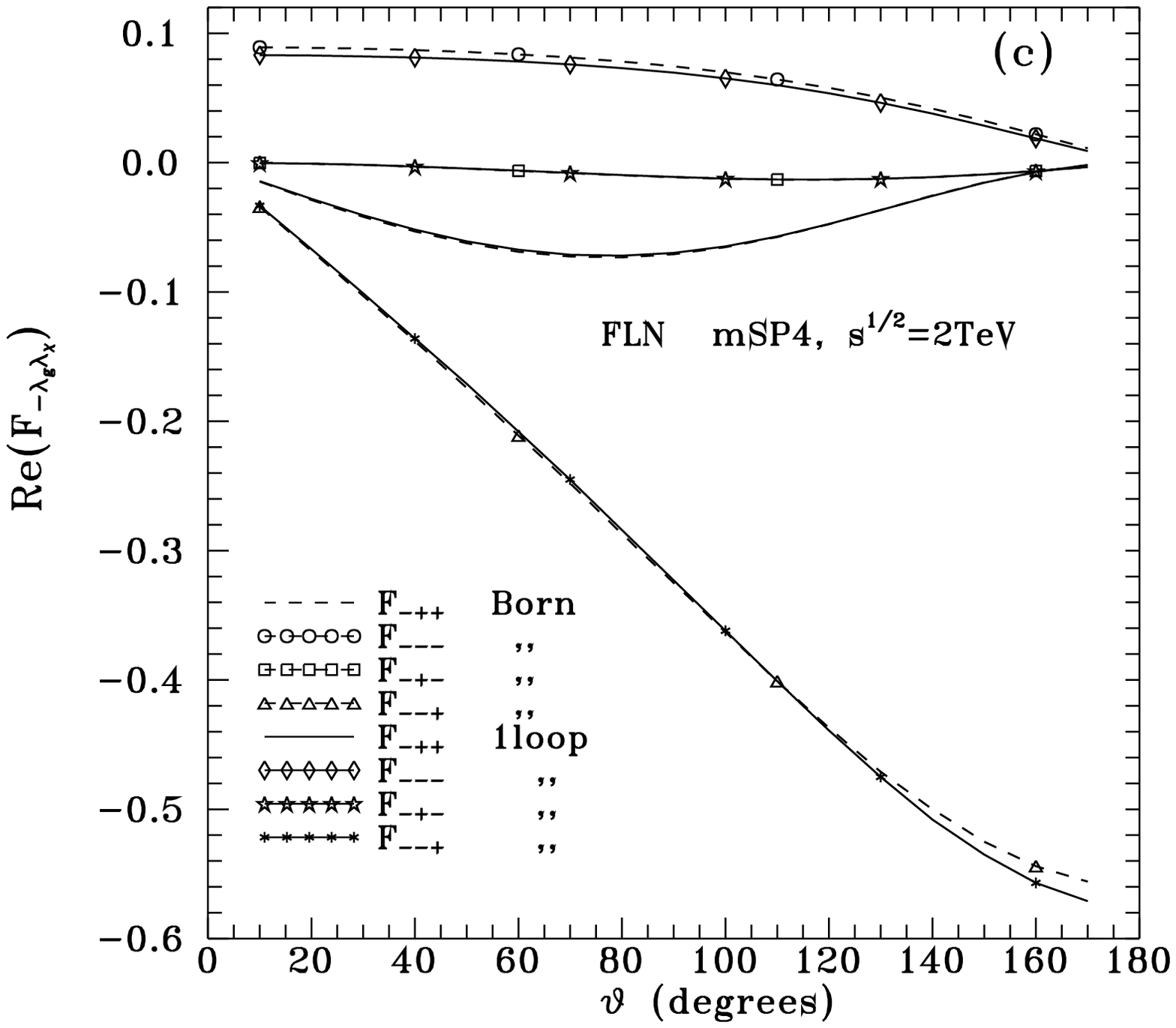,height=7.5cm}\hspace{0.5cm}
\epsfig{file=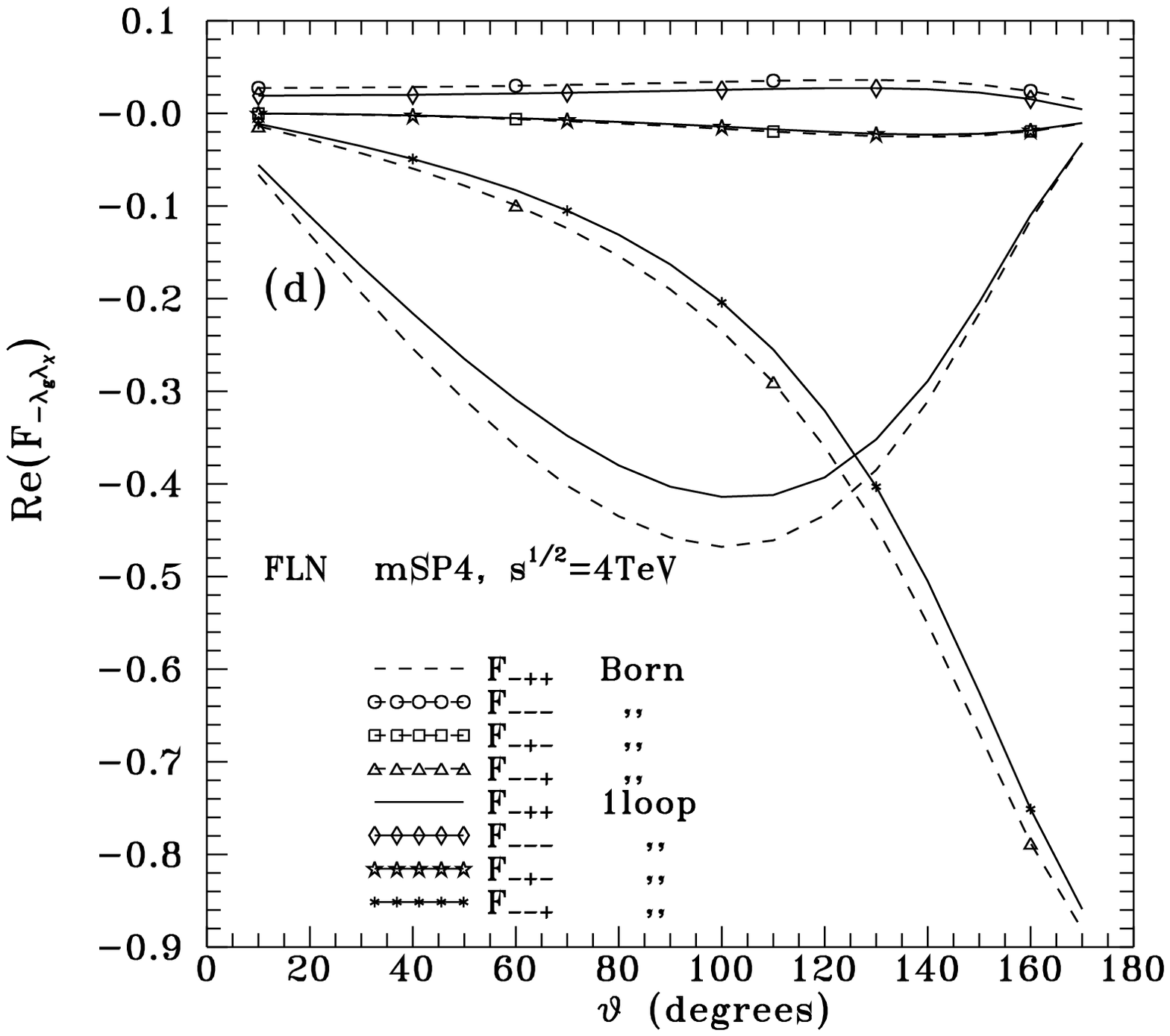,height=7.5cm}
\]
\caption[1]{The  $ug\to \tilde d_L \tchi^+_1$ helicity amplitudes
for the FLN mSP4 model. The  energy dependencies
cover an LHC-type range (a), and a higher energy region (b);
while the  angular dependencies are given
at $\sqrt{s}=2$TeV (c), and $\sqrt{s}=4$TeV (d).  }
\label{Sp4-amp-fig}
\end{figure}

\begin{figure}[p]
\vspace*{-2.cm}
\[
\hspace{-0.5cm}
\epsfig{file=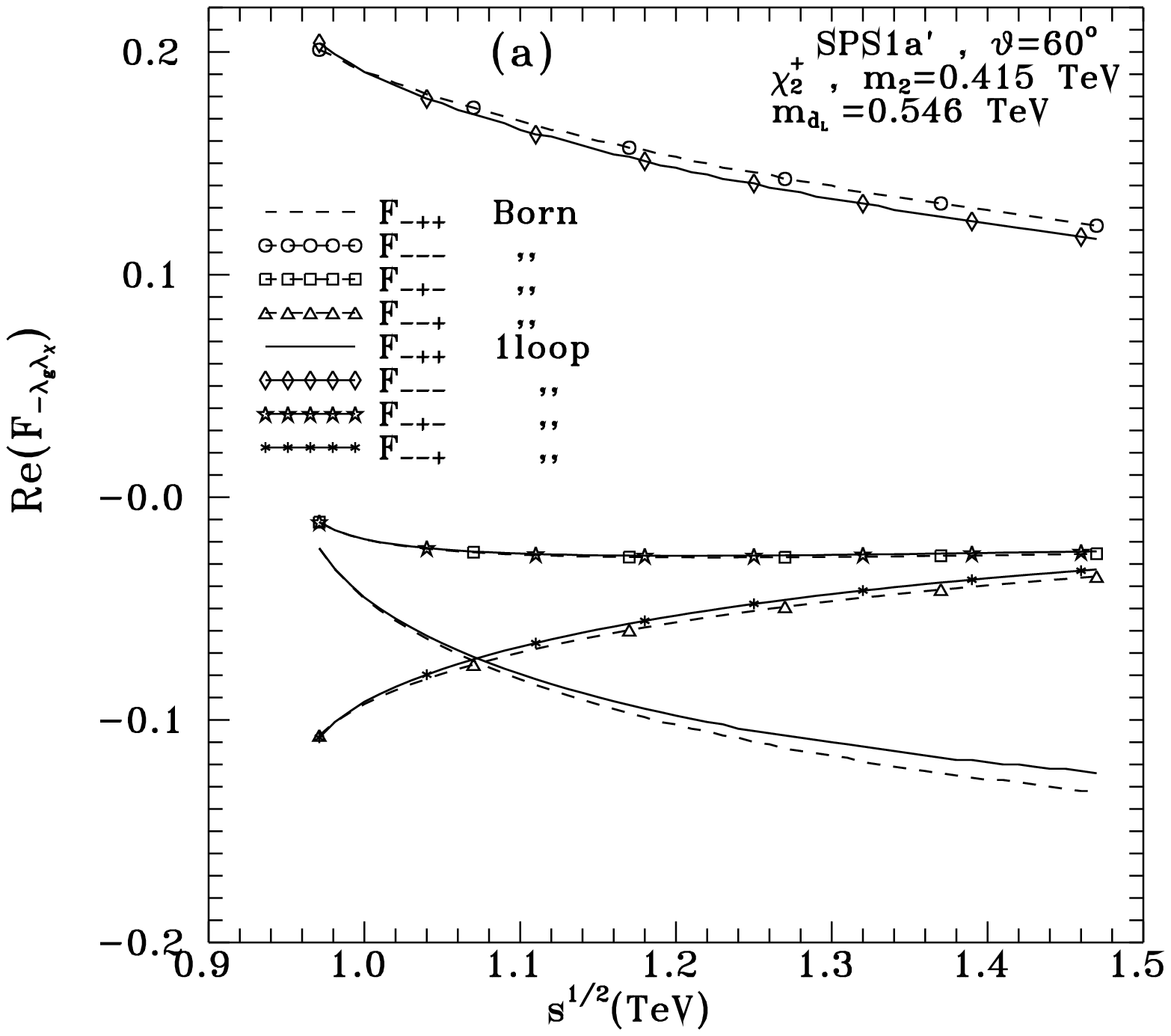,height=7.5cm}\hspace{0.5cm}
\epsfig{file=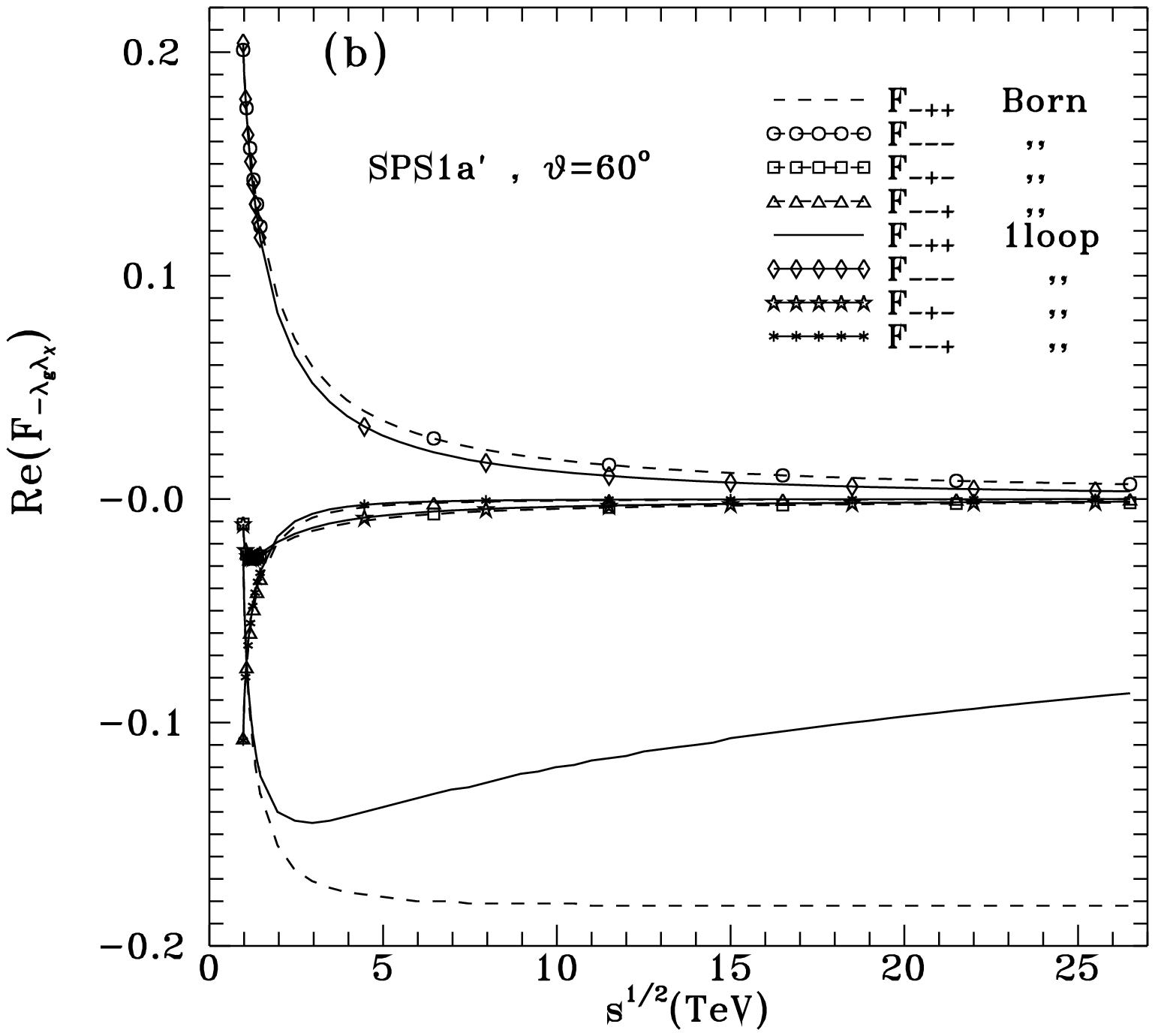,height=7.5cm}
\]
\vspace*{0.5cm}
\[
\hspace{-0.5cm}
\epsfig{file=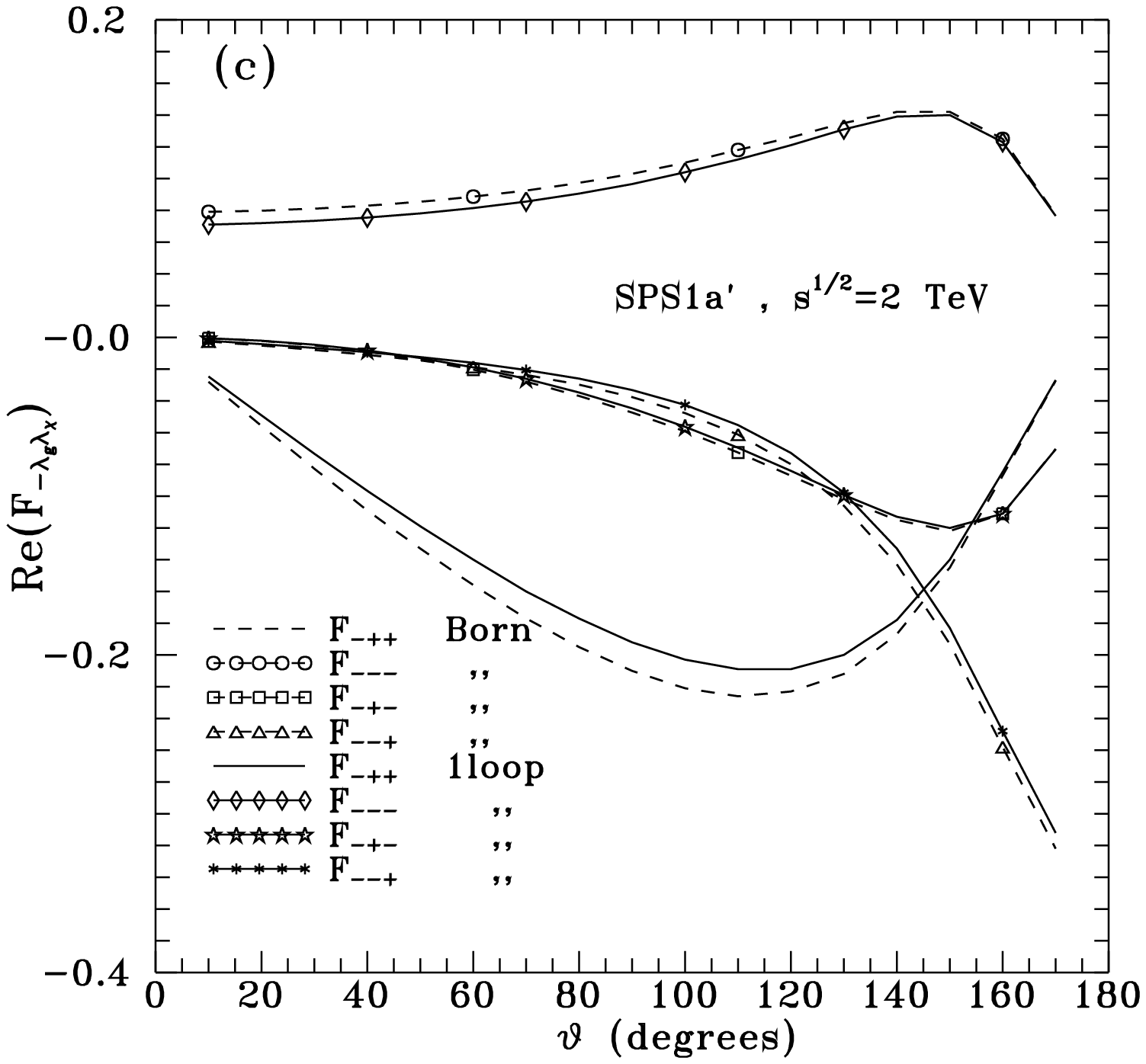,height=7.5cm}\hspace{0.5cm}
\epsfig{file=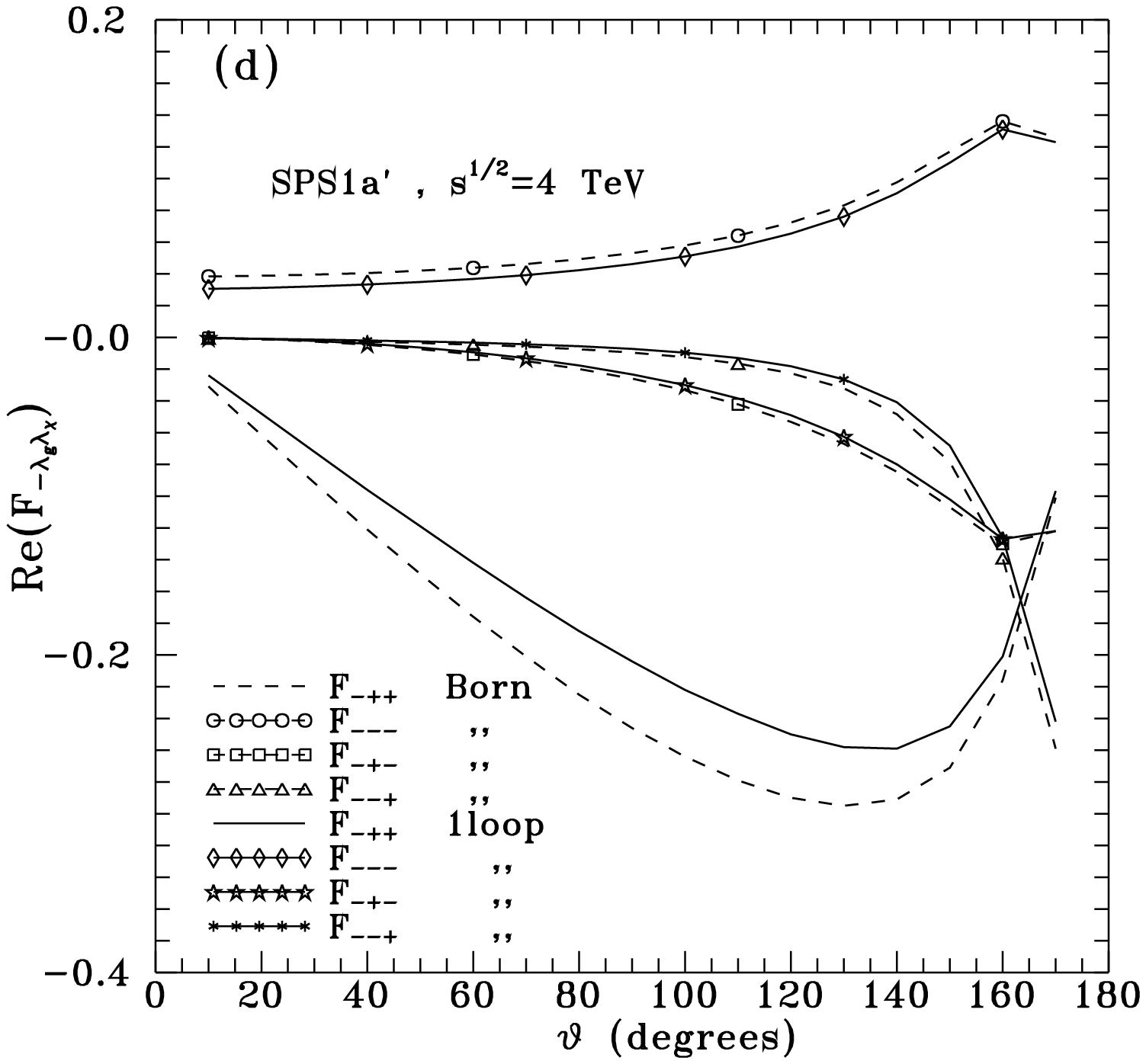,height=7.5cm}
\]
\caption[1]{As in Fig.\ref{SPA-amp-fig}, but for
the  $ug\to \tilde d_L \tchi^+_2$ amplitudes. The  energy dependencies
cover an LHC-type range (a), and a higher energy region (b);
while the  angular dependencies are given
at $\sqrt{s}=2$TeV (c), and $\sqrt{s}=4$TeV (d).  }
\label{SPA-amp2-fig}
\end{figure}

\begin{figure}[p]
\vspace*{-2.cm}
\[
\hspace{-0.5cm}
\epsfig{file=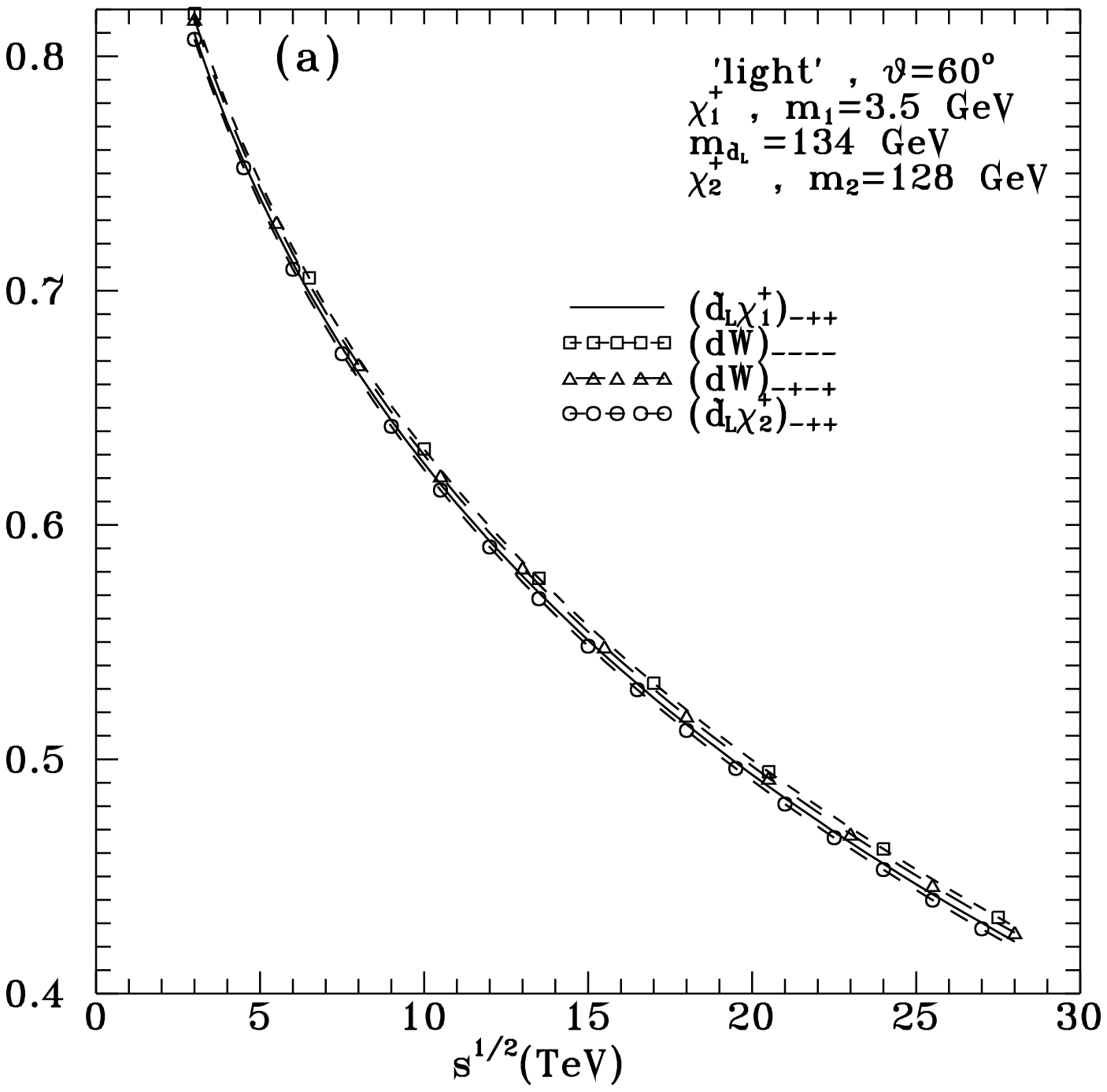,height=7.5cm}\hspace{0.5cm}
\epsfig{file=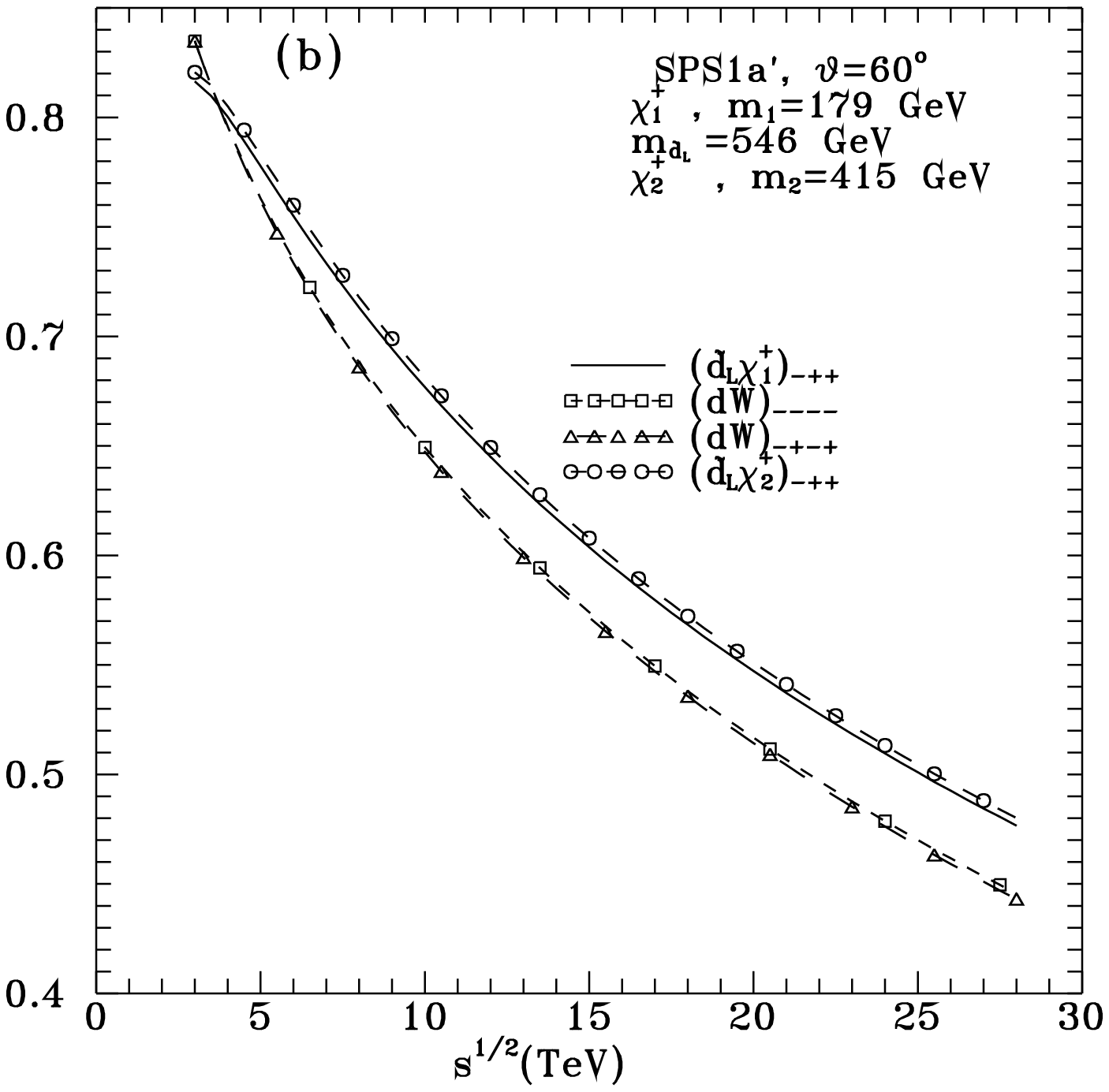,height=7.5cm}
\]
\vspace*{0.5cm}
\[
\hspace{-0.5cm}
\epsfig{file=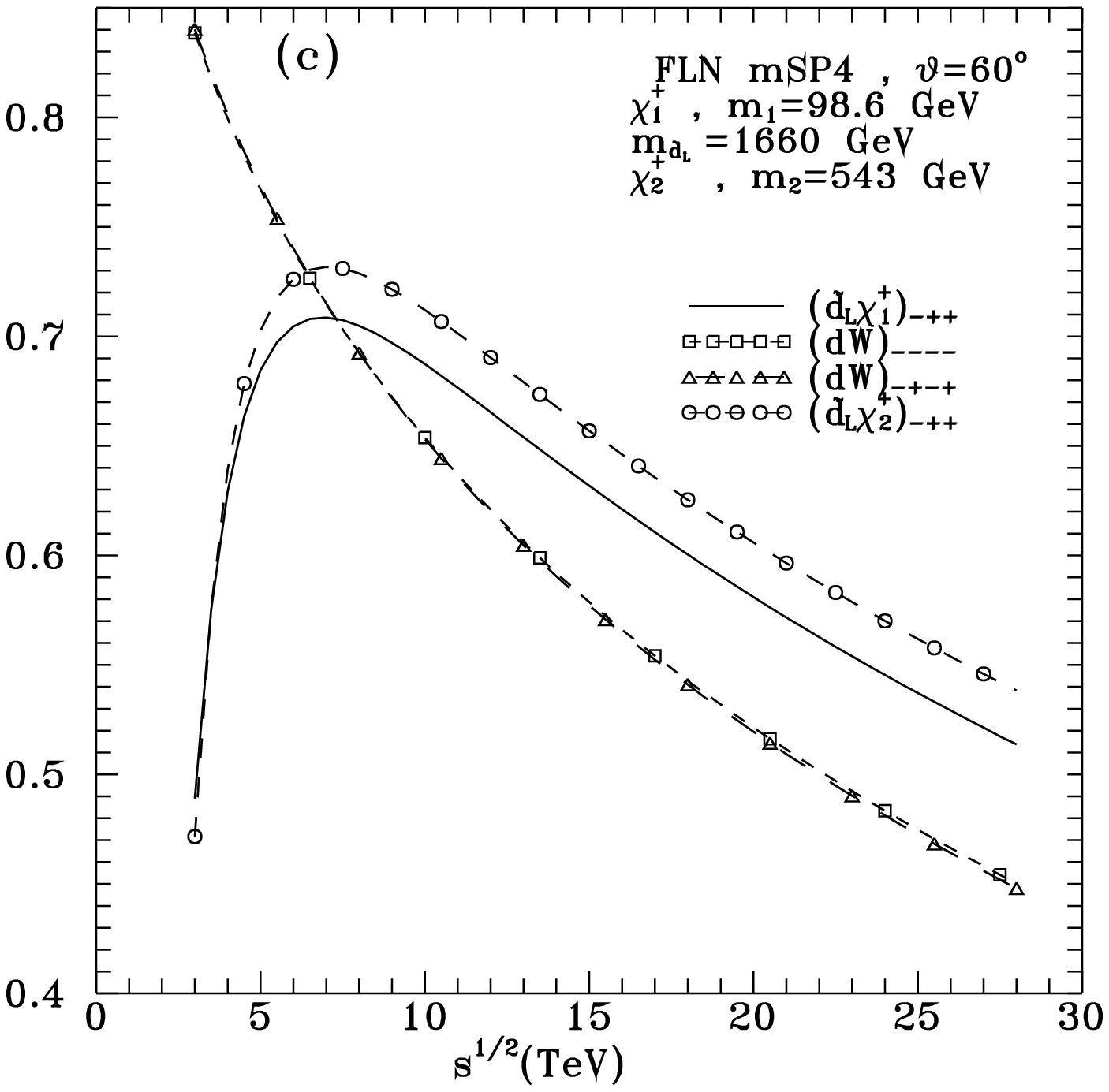,height=7.5cm}
\]
\caption[1]{The energy dependencies of  the $W$-production  parts
of (\ref{test-F-relation}), are compared to those for $\tchi^+_1$ (full line) and
 $\tchi^+_2$  (dash  line with circles) production,
 using    the models "light" (a), $SPS1a'$ (b), and
FLN mSP4 (c). The scattering angle is fixed at $\theta=60^o$.
The corresponding masses of  $\tchi^+_1$,  $\tchi^+_2$ and $\tilde d_L$
are also indicated. }
\label{F-relation-fig}
\end{figure}

\begin{figure}[p]
\vspace*{-2.cm}
\[
\hspace{-0.5cm}
\epsfig{file=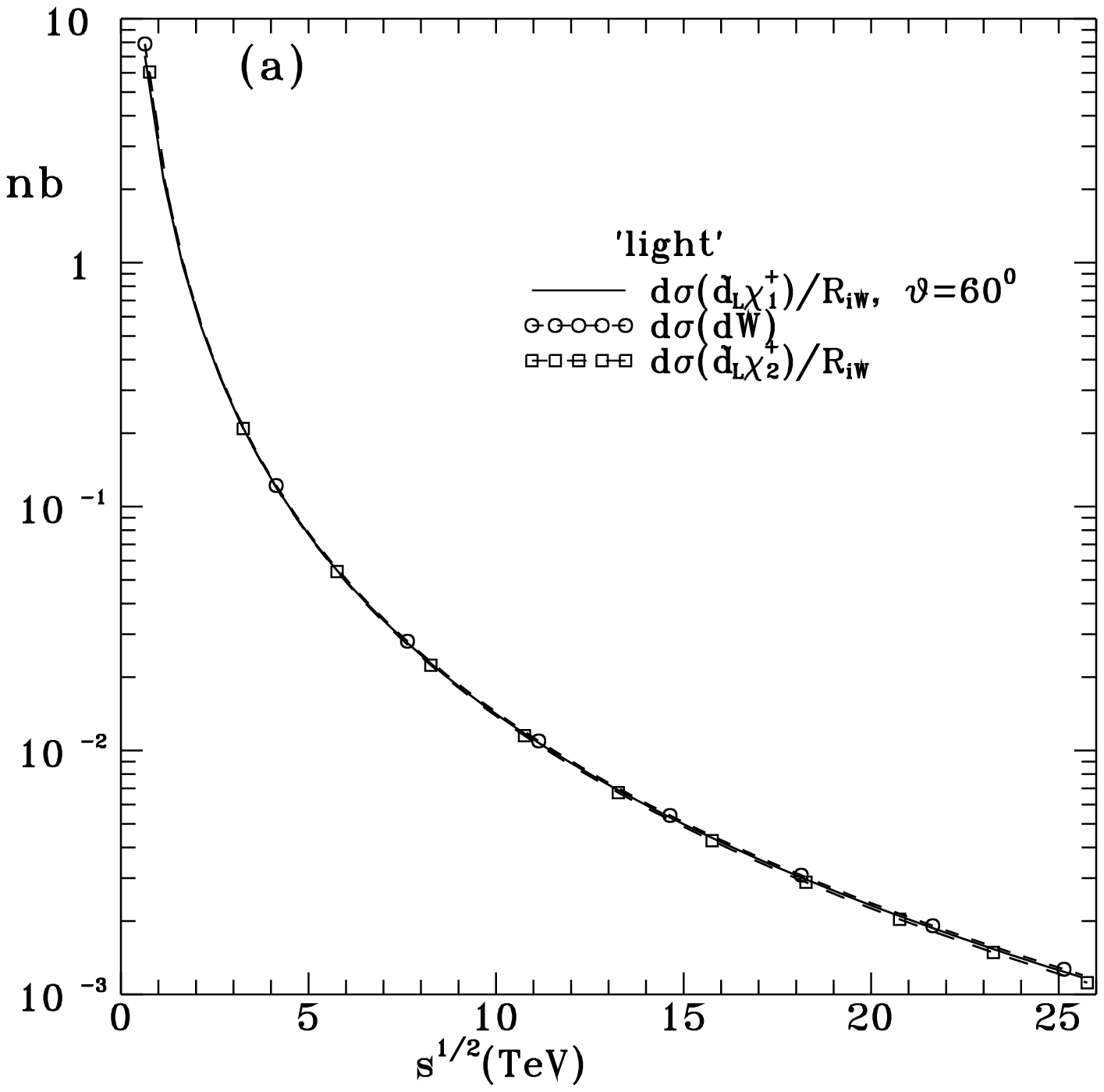,height=7.5cm}\hspace{0.5cm}
\epsfig{file=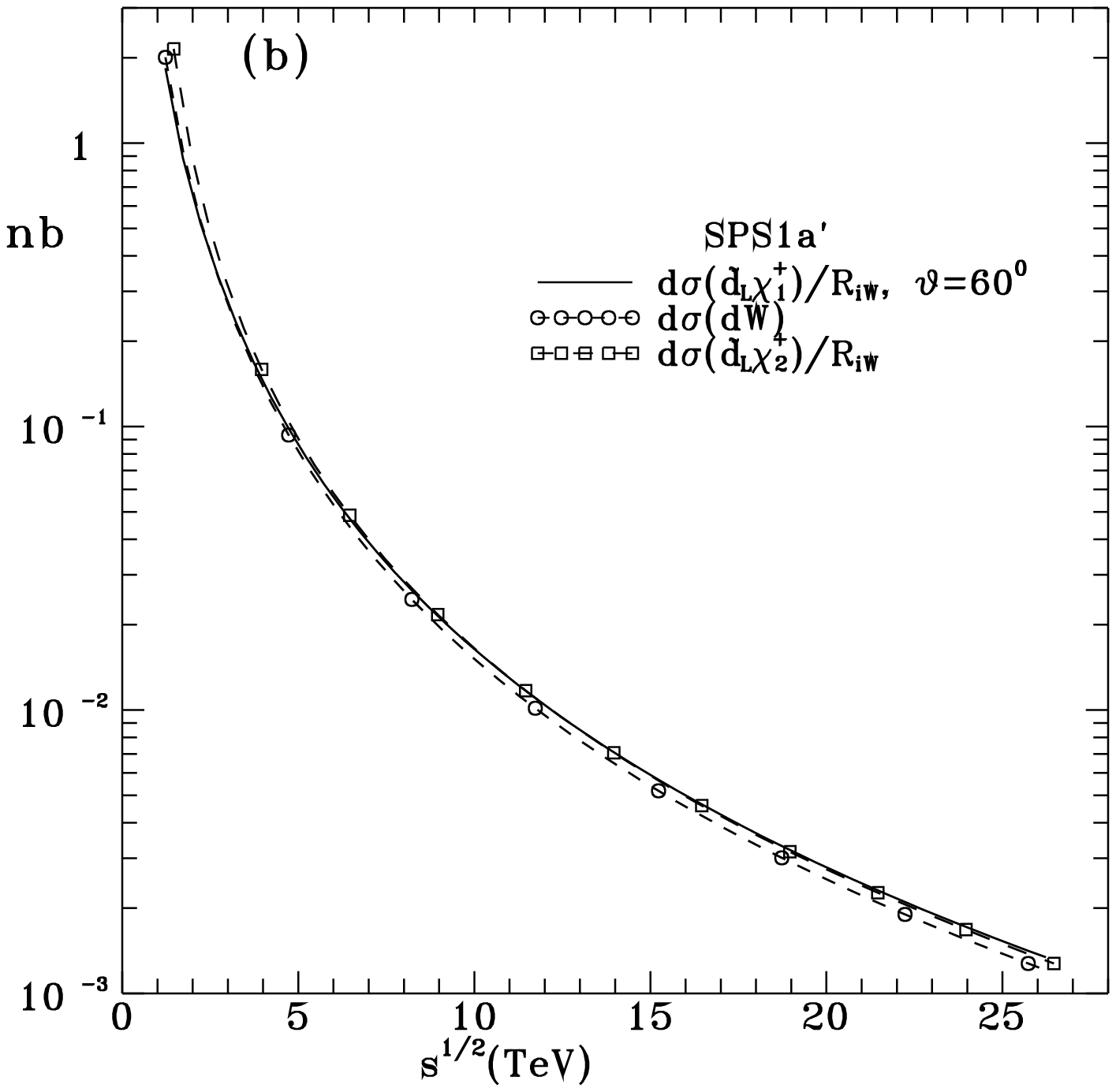,height=7.5cm}
\]
\vspace*{0.5cm}
\[
\hspace{-0.5cm}
\epsfig{file=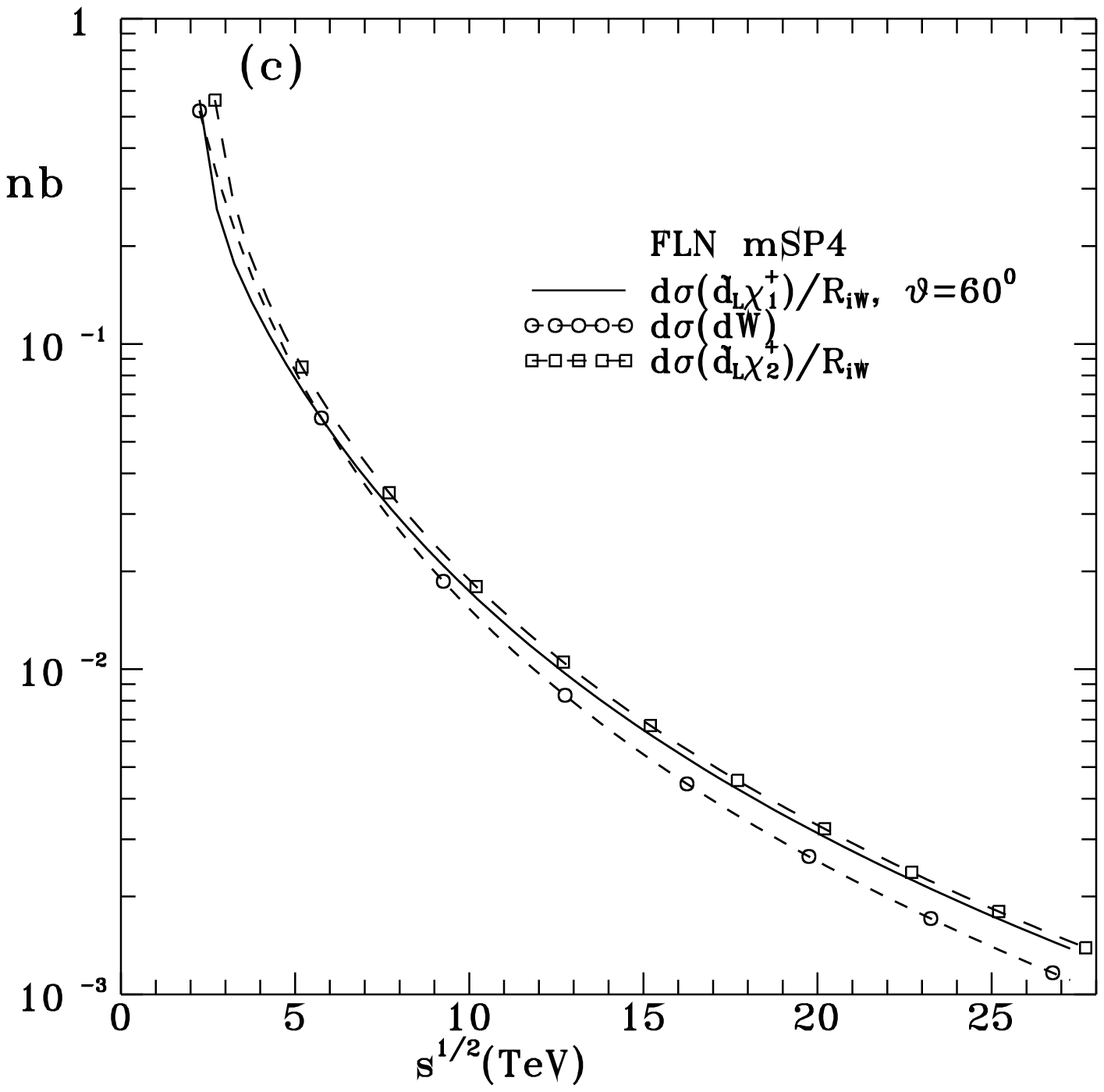,height=7.5cm}
\]
\caption[1]{ The energy dependencies of the left part
of (\ref{test-sigma-relation}) (dash line with circles), are compared
to the    $\tchi^+_1$ (full line)
and  $\tchi^+_2$  (dash line with squares) production parts,
  using  the models "light" (a), $SPS1a'$ (b), and
FLN mSP4 (c). The scattering angle is fixed at $\theta=60^o$.
The corresponding masses of  $\tchi^+_1$,  $\tchi^+_2$ and $\tilde d_L$
 are given in Fig.\ref{F-relation-fig}.}
\label{Signa-e-relation-fig}
\end{figure}

\begin{figure}[p]
\vspace*{-2.cm}
\[
\hspace{-0.5cm}
\epsfig{file=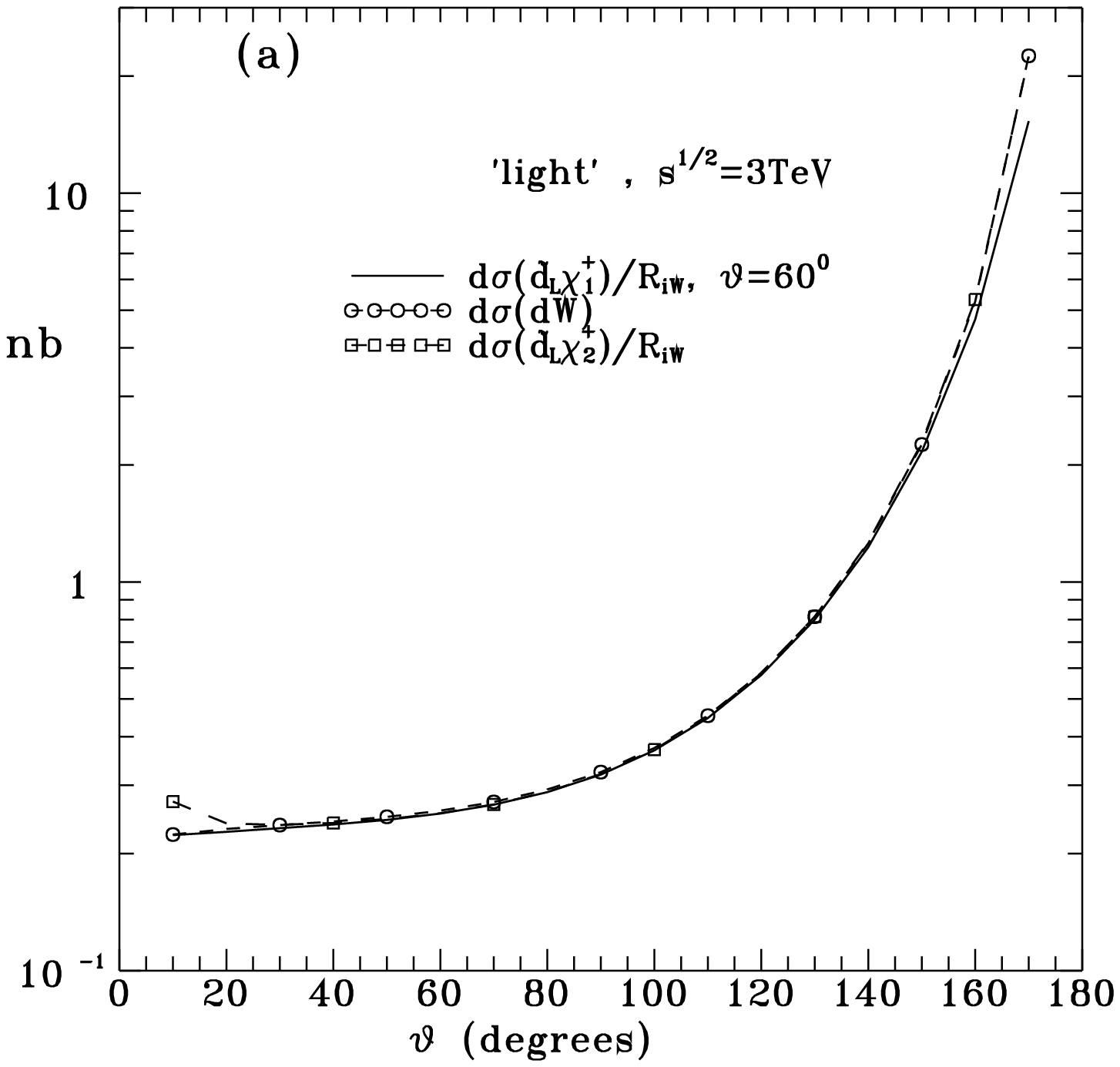,height=7.5cm}\hspace{0.5cm}
\epsfig{file=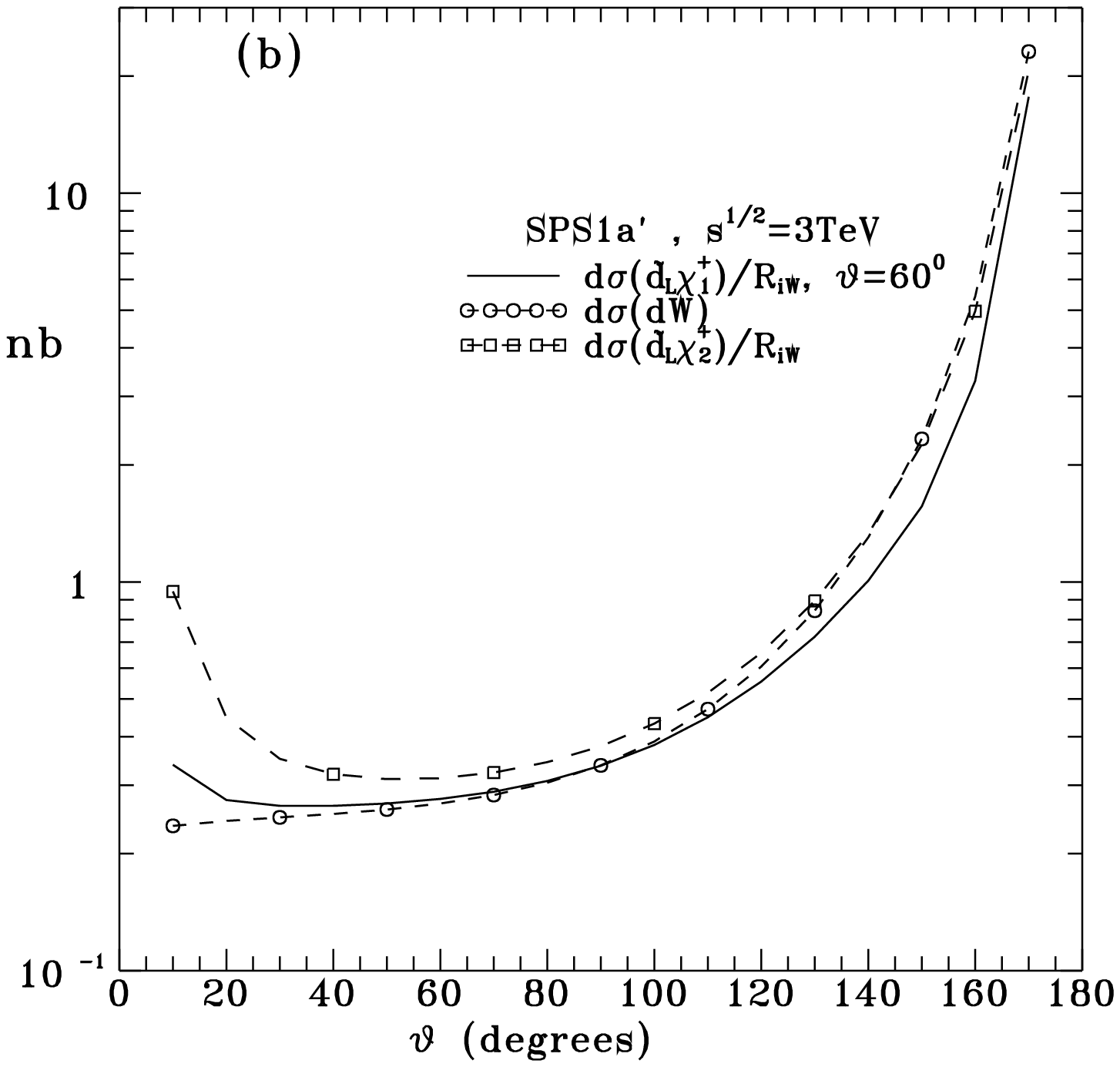,height=7.5cm}
\]
\vspace*{0.5cm}
\[
\hspace{-0.5cm}
\epsfig{file=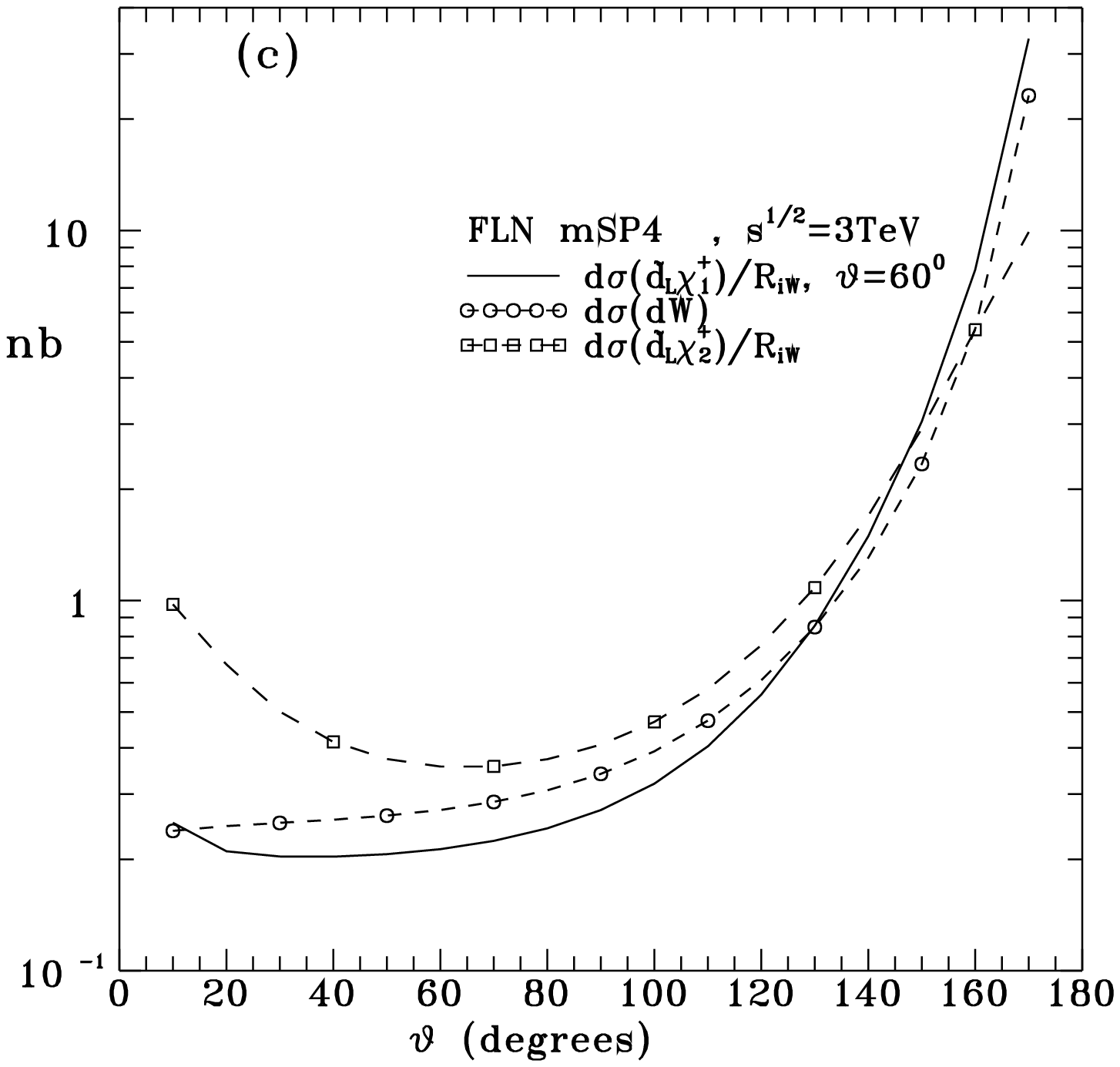,height=7.5cm}
\]
\caption[1]{The  angular  dependencies of the left part
of (\ref{test-sigma-relation}) (dash line with circles), are compared
to the  $\tchi^+_1$ (full line)
and  $\tchi^+_2$ (dash line with squares) production parts,
  using  the models "light" (a), $SPS1a'$ (b), and
FLN mSP4 (c). The energy is fixed  at $\sqrt{s}=3$TeV.
The corresponding masses of  $\tchi^+_1$,  $\tchi^+_2$ and $\tilde d_L$
 are given in Fig.\ref{F-relation-fig}. }
\label{Signa-a-relation-fig}
\end{figure}

\begin{figure}[p]
\vspace*{-2.cm}
\[
\hspace{-0.5cm}
\epsfig{file=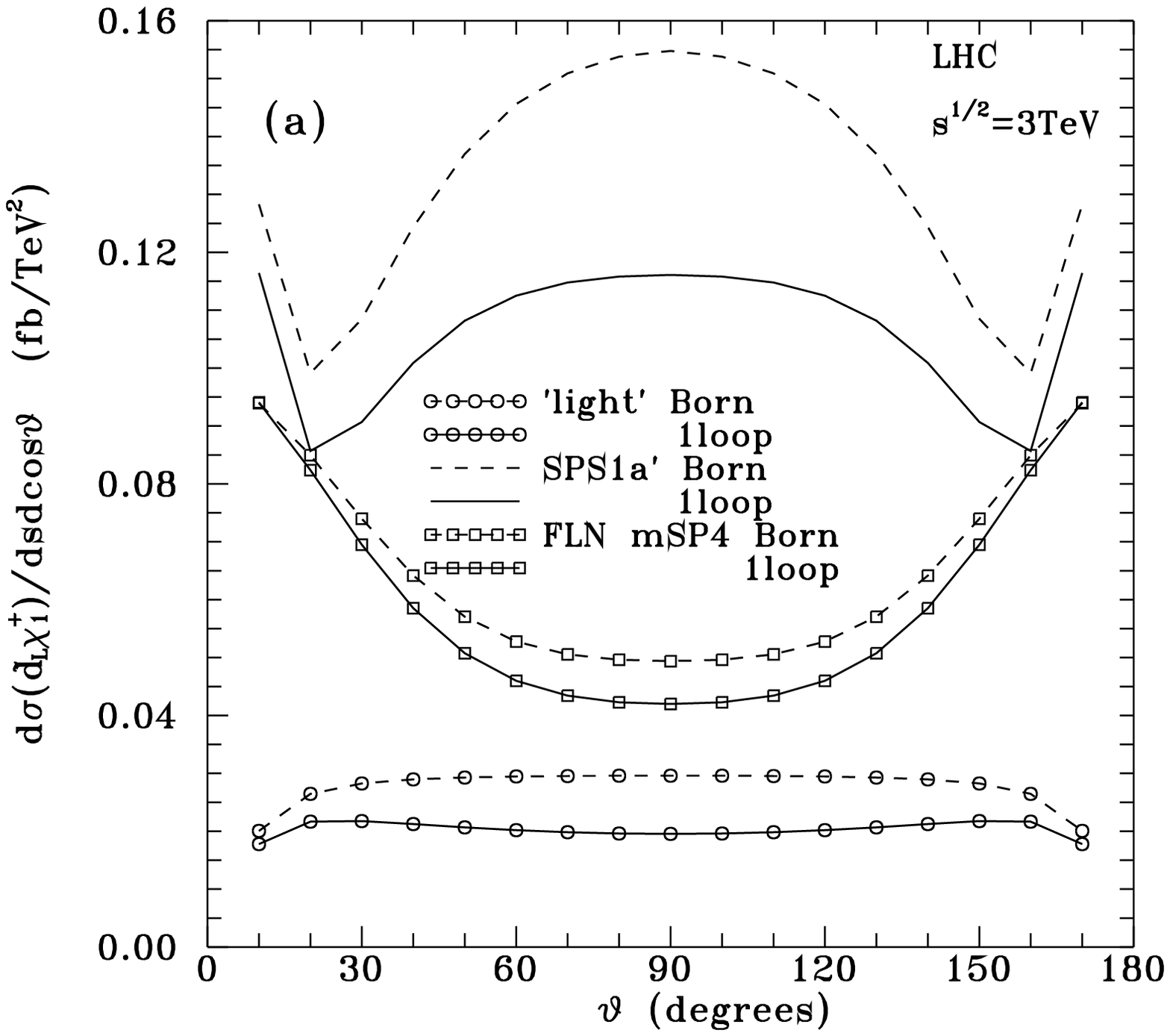,height=7.5cm}\hspace{0.5cm}
\epsfig{file=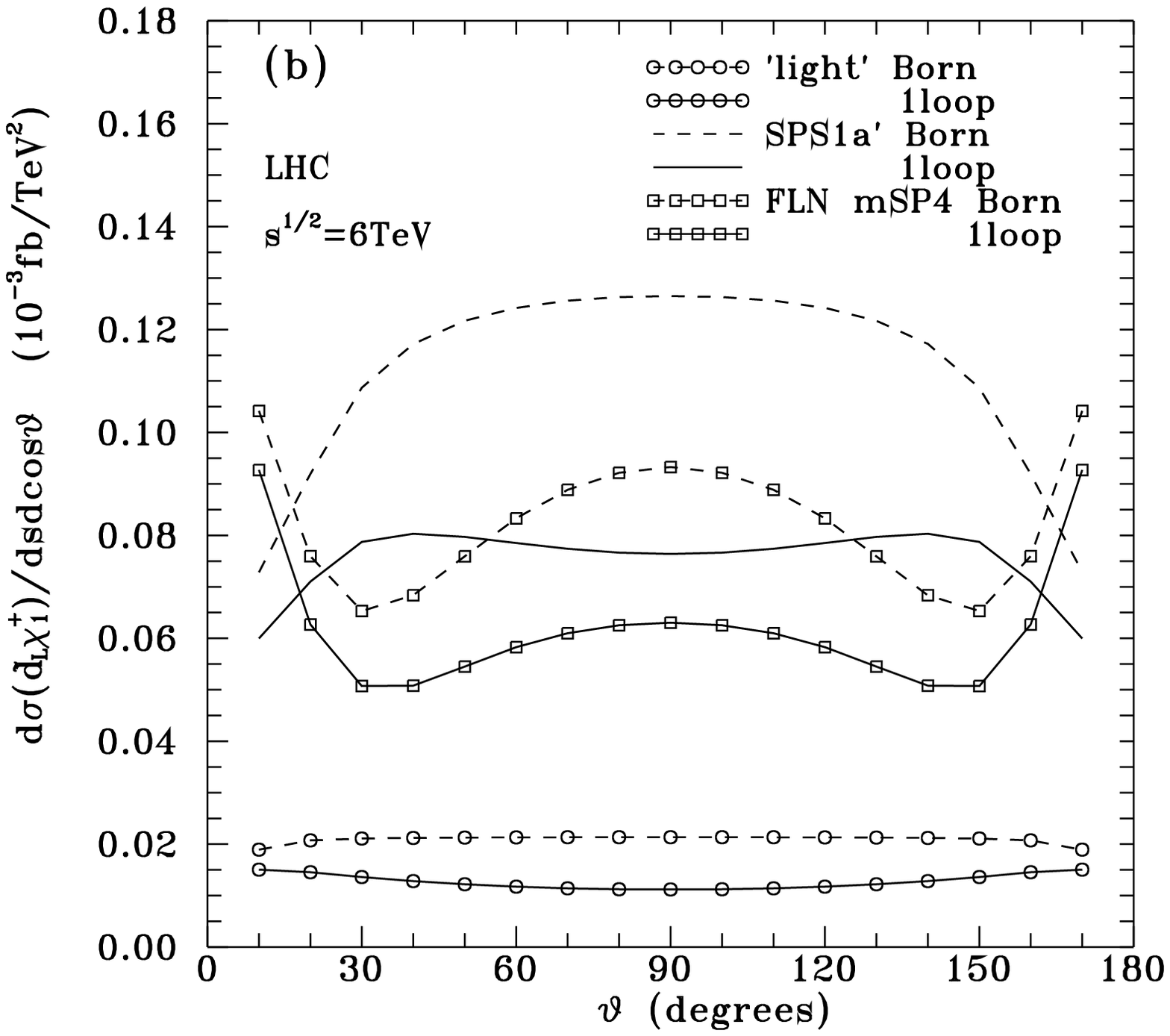,height=7.5cm}
\]
\vspace*{0.5cm}
\[
\hspace{-0.5cm}
\epsfig{file=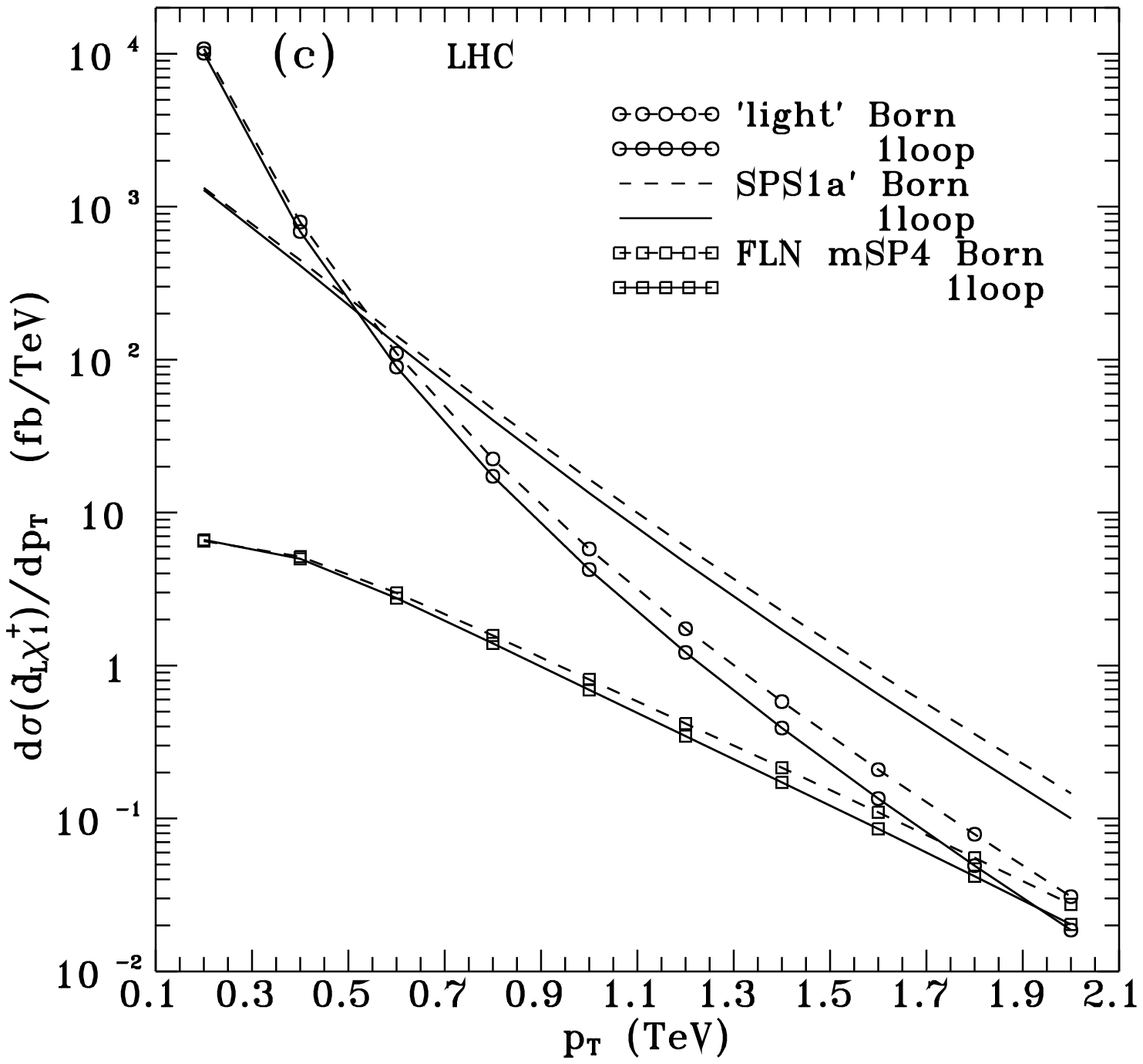,height=7.5cm}
\]
\caption[1]{Born (dash line) and 1loop (full line) predictions for the
LHC  distributions in  $\tilde d_L \tchi^+_1$ production, for  the
3 models of Table 1; (a) gives the angular distributions
at $\sqrt{s}=3$TeV, (b) the angular distribution at   $\sqrt{s}=6$TeV,
 (c) the  transverse momentum  distributions. }
\label{LHC-fig}
\end{figure}

The first aim  is to  understand
how the asymptotic Helicity Conservation property
for     $ug\to \tilde d_L \tchi_i^+$, reflects itself, as the energy is reduced
to   non-asymptotic or even LHC values. As compared to the
$ug\to dW$ case,  the establishment of HC in  $\tilde d_L \tchi_i^+$-production
is delayed, by the higher masses of the produced particles.
But HC may nevertheless be visible at subprocess c.m. energies of about 4TeV,
if the SUSY scale is not too high. Compare the results in Fig.\ref{SPA-amp-fig}
and  \ref{SPA-amp2-fig}
for $\tilde d_L \tchi_1^+$ and  $\tilde d_L \tchi_2^+$ production
respectively, based on the $SPS1a'$-model \cite{SPA}.
The recent  very  precise analysis
of \cite{OBuchmueller, Nath}, allows entertaining the  hope
that this is a viable possibility in Nature.

The second aim  concerns at identifying simple SUSY relations between
the processes $ug\to \tilde d_L \tchi_i^+$ and $ug\to dW^+$, characterized by
the same initial state, but having  their final states constituting
supersymmetric particle pairs. Assuming energies much higher than
 all SUSY masses, we derive two such relations affecting the dominant
high energy amplitudes and the corresponding cross sections, called
respectively  {\it F-relation} and $\sigma$-{\it relation} respectively.
Using then three model examples covering a reasonable scale
of SUSY scales, we investigate  how the deviations of these relations develop,
as  the energy is reduced down to the LHC range.
 Particularly for the $\sigma$-{\it relation}, we have found
 that it  may be quite accurate at LHC energies, or so;
 provided the SUSY scale is not much larger than the one of the
  FLN mSP4 model of Table 1.   If this is case,
  they may be used in testing the consistency of identifying
  a pair two new  particles  produced  at LHC, as
  consisting of a $\tilde d_L$ and a chargino.
  This seems even more true for the $\tchi^+_1$ case,
  probably due to the lighter chargino mass.    \\

The third aim was to present the Born contribution and the
1loop EW corrections, to the   $\tilde d_L \tchi^+_i$ production
at LHC, without any  high energy assumptions. Both,
the angular and transverse momentum distributions were studied.
As in the $W$+jet production case \cite{ugdW-our}, the SUSY 1loop corrections
were always found to reduce the Born contribution, roughly by an amount
at the 10\% level. This may be  observable,
provided the SUSY scale is not too high.

Combining in fact this  $\tilde d_L \tchi^+_i$ production study,
with the corresponding
one for $W$+jet production \cite{ugdW-our}, offers  stringent tests  of
the nature of candidate  supersymmetric particles. Because,
we should not only have a reasonable magnitude for the Born contribution
to $ug\to \tilde d_L \tchi^+_i $, but also the 1loop EW corrections
to this process, as well as to $ug\to dW$, should come out right.\\

Finally, we come  back to the intriguing
helicity conservation property of any  2-to-2 body process
at asymptotic energies, in a softly broken  renormalizable
supersymmetric
theory\footnote{All anomalous couplings we are aware of, violate HC
\cite{Kasimierz}. } \cite{heli}. Its  realization comes about
after the appearance of huge  cancellations among the various diagrams. Both,
here and in previous work \cite{ugdW-our, ggVV},
we were  fascinated to see this happening in detail,
so  that no terms involving ratios of masses destroy it. This is most tricky,
when longitudinal gauge bosons and Yukawa couplings
are involved;  we intend to examine such cases in the future.

Of course, since HC is an asymptotic theorem, its phenomenological
relevance depends mainly on the external masses. If the external masses
are not too heavy, like in $ug\to dW$, it may be partly
realized already at LHC energies \cite{ugdW-our}. If the masses are heavier,
like in the present $ug \to \tilde d_L \tchi^+_i$ example,
its realization is delayed. In any case though, it
provides a stringent test of any
theoretical calculation of such supersymmetric processes .\\

\vspace*{1cm}
\noindent
{\large\bf{Acknowledgements}}\\
\noindent
G.J.G. gratefully acknowledges the support by the European Union  contracts
MRTN-CT-2004-503369 and   HEPTOOLS,  MRTN-CT-2006-035505.

\newpage

\newpage

\newpage

\newpage

\newpage

\newpage

\newpage

\newpage

\newpage

\end{document}